\RequirePackage{lineno}
\documentclass[prd,amsmath,amssymb,showpacs,superscriptaddress,nofootinbib,twocolumn]{revtex4-1}
\usepackage{graphicx}
\usepackage{epsfig}
\usepackage{dcolumn}
\usepackage{bm}
\usepackage{overpic}
\usepackage{color}  
\usepackage[colorlinks,linkcolor=blue,anchorcolor=blue,citecolor=blue,dvipdfm]{hyperref} 

\begin{document}
\normalsize

\parskip=5pt plus 1pt minus 1pt

\title{\boldmath Measurement of the proton form factor by studying $e^{+} e^{-}\rightarrow p\bar{p}$}

\author{
  \begin{small}
    \begin{center}
      M.~Ablikim$^{1}$, M.~N.~Achasov$^{9,a}$, X.~C.~Ai$^{1}$,
      O.~Albayrak$^{5}$, M.~Albrecht$^{4}$, D.~J.~Ambrose$^{44}$,
      A.~Amoroso$^{48A,48C}$, F.~F.~An$^{1}$, Q.~An$^{45}$,
      J.~Z.~Bai$^{1}$
% ...
      \\
      \vspace{0.2cm}
      (BESIII Collaboration)\\
      \vspace{0.2cm} {\it
        $^{1}$ Institute of High Energy Physics, Beijing 100049, People's Republic of China\\
% ...
$^{4}$ Bochum Ruhr-University, D-44780 Bochum, Germany\\
        $^{5}$ Carnegie Mellon University, Pittsburgh, Pennsylvania 15213, USA\\
% ...
$^{9}$ G.I. Budker Institute of Nuclear Physics SB RAS (BINP), Novosibirsk 630090, Russia\\
% ...
$^{44}$ University of Rochester, Rochester, New York 14627, USA\\
        $^{45}$ University of Science and Technology of China, Hefei 230026, People's Republic of China\\
% ...
$^{48}$ (A)University of Turin, I-10125, Turin, Italy; (B)University of Eastern Piedmont, I-15121, Alessandria, Italy; (C)INFN, I-10125, Turin, Italy\\
% ...
$^{a}$ Also at the Novosibirsk State University, Novosibirsk, 630090, Russia\\
% ...
      }\end{center}
    \vspace{0.4cm}
  \end{small}
}

\affiliation{}

\date{\today}

\begin{abstract}
Using data samples collected with the BESIII detector at the BEPCII collider, we measure the
Born cross section of $e^{+}e^{-}\rightarrow p\bar{p}$ at 12 center-of-mass energies
from 2232.4 to 3671.0 MeV.
The corresponding effective electromagnetic form factor of the proton is deduced under the
assumption that the electric and magnetic form factors are equal $(|G_{E}|= |G_{M}|)$. In addition,
the ratio of electric to magnetic form factors, $|G_{E}/G_{M}|$, and $|G_{M}|$ are
extracted by fitting the polar angle distribution of the proton for the data samples with larger
statistics, namely at $\sqrt{s}=$ 2232.4 and 2400.0 MeV and a combined sample at $\sqrt{s}$ = 3050.0,
3060.0 and 3080.0 MeV, respectively.
The measured cross sections are in agreement with recent results from BaBar, improving the overall
uncertainty by about 30\%. The $|G_{E}/G_{M}|$ ratios are close to unity
and consistent with BaBar results in the same $q^{2}$ region, which indicates the data are
consistent with the assumption that $|G_{E}|=|G_{M}|$ within uncertainties.
\end{abstract}

\maketitle

\section{\boldmath Introduction}

Electromagnetic form factors (FFs) of the nucleon
provide fundamental information about its internal structure and dynamics.
They constitute a rigorous test of non-perturbative QCD as well as of phenomenological
models.

\begin{figure*}[htbp]
  \centering
  \begin{overpic}[width=0.48\textwidth]{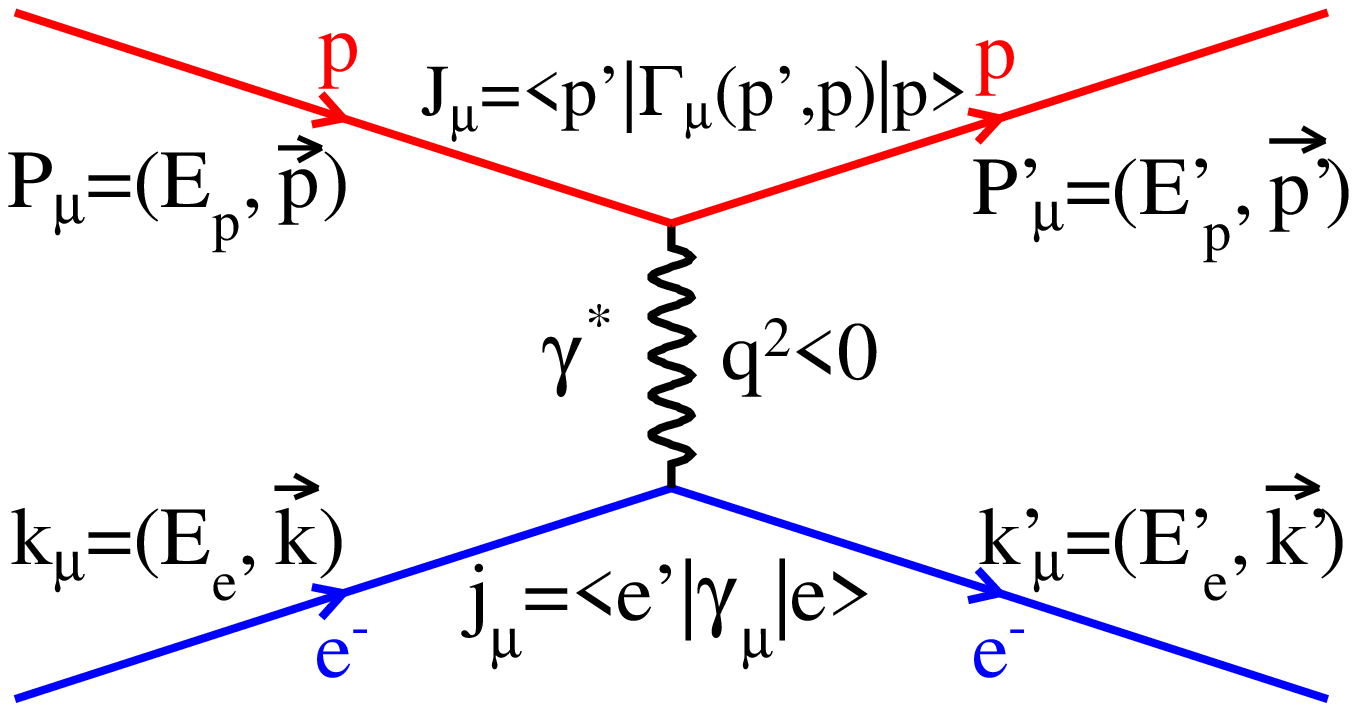}
    \put(45,45){ (a)}
  \end{overpic}
  \begin{overpic}[width=0.48\textwidth]{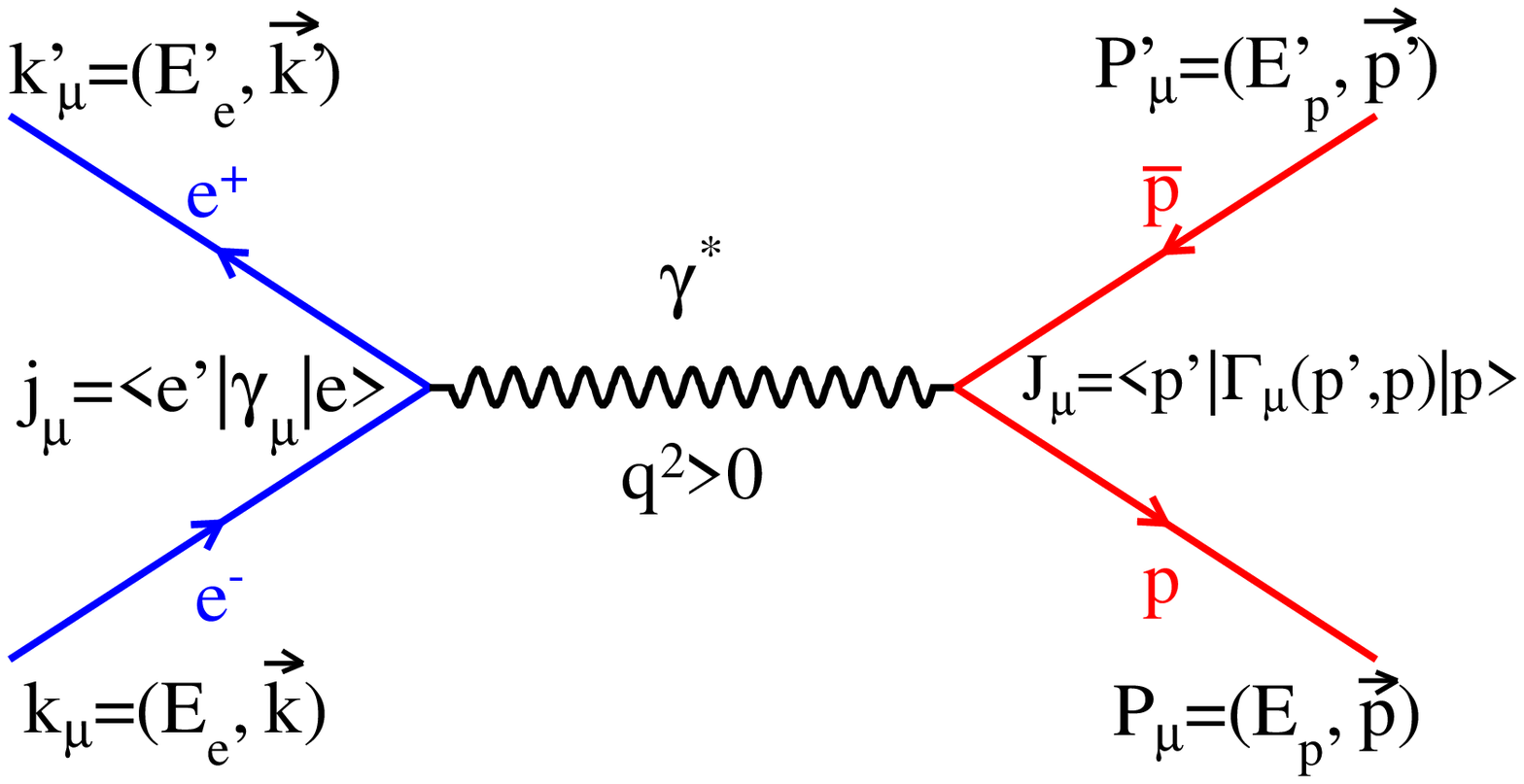}
    \put(45,45){ (b)}
  \end{overpic}
  \vskip -0.3cm
  \parbox[1cm]{16cm} {
    \caption{(a) Feynman diagram of $e p\rightarrow e p$ elastic scattering at the lowest order.
      (b) Feynman diagram of $e^{+}e^{-}\rightarrow p\bar{p}$ annihilation at the lowest order
      (identical to that of the reverse reaction $p\bar{p}\rightarrow e^{+}e^{-}$ with $e\leftrightarrow p$ exchange.)}
    \label{feynman1}
  }
\end{figure*}

Proton FFs can be measured in different kinematic regions by {\it i)} lepton-proton elastic
scattering (space-like, labeled SL) {\it ii)} electron-positron annihilation into a
proton-antiproton pair or proton-antiproton annihilation into an electron-positron (time-like,
labeled TL).  The lowest order Feynman diagram of lepton-proton scattering is shown in
Fig.~\ref{feynman1}(a). The momentum transfer squared, $q^{2}$, is negative and the FFs are real
functions of $q^{2}$.  The lowest order $e^{+}e^{-}$ annihilation process is shown in
Fig.~\ref{feynman1}(b). Here, $q^{2}$ is positive and the FFs are complex functions of $q^{2}$.  The
basic kinematic variables are also shown in Fig.~\ref{feynman1}, where $k$, $k'$ are the
initial and final electron momenta and $p$, $p'$ are the initial and final proton momenta.
Since the electromagnetic vertex of the lepton is well known, one can reliably extract the proton
electromagnetic vertex $\Gamma^{\mu}$ by measuring the cross section and the polarization. Assuming
one-photon exchange, i.e. in Born approximation, and under the basic requirements of Lorentz
invariance, the hadronic vertex can be parameterized in terms of two FFs, $F_{1}$ and
$F_{2}$~\cite{ff},
\begin{equation}
  \Gamma_{\mu}(p',p)= \gamma_{\mu} F_{1}(q^{2})+\frac{i\sigma_{\mu\nu}q^{\nu}}{2m_{p}}\kappa_{p}F_{2}(q^{2}),
  \label{f1}
\end{equation}
where the element $\sigma_{\mu\nu}=\gamma_{\mu}\gamma_{\nu}-\gamma_{\nu}\gamma_{\mu}$ is a
representation of the Lorentz group,
$m_{p}$ is the mass of the proton, $\kappa_{p}=\frac{g_{p}-2}{2}$ is the anomalous magnetic moment
of the proton, $g_{p}=\frac{\mu_{p}}{J}$, $\mu_{p}=2.79$ is the magnetic moment of the proton and
$J=\frac{1}{2}$ is its spin.
The functions $F_{1}$ and $F_{2}$ are the so called Dirac and Pauli FFs, respectively.  The optical
theorem, applied to lepton-nucleon scattering, implies that at the lowest order the FFs are real in
the SL region~\cite{optical}~\cite{optical2}, i.e. the complex conjugate of the amplitude in Fig.~\ref{feynman1}(a),
$\mathcal{M}^{\mathcal{\dagger}}$, is identical to $\mathcal{M}$. In the TL region, as in in
Fig.~\ref{feynman1}(b), the FFs can be complex above the first hadronic threshold, that is, above
twice the pion mass.

The Sachs FFs, electric $G_{E}$ and magnetic $G_{M}$, are introduced
as linear combinations of the Dirac
and Pauli FFs~\cite{sachs}. Concerning the SL region in the Breit frame, $G_{E}$ and $G_{M}$ are the Fourier
transforms of the charge and magnetization distribution of the nucleon, respectively. $G_{M}$ and
$G_{E}$ are proportional to spin-flip and non spin-flip amplitudes, respectively.  They are
expressed as
\begin{equation}
  G_{E}(q^{2}) = F_{1}(q^{2}) + \frac{q^{2}}{4m_{p}^{2}}\kappa_{p}F_{2}(q^{2}),
  \label{f2}
\end{equation}
\begin{equation}
  G_{M}(q^{2}) = F_{1}(q^{2}) + \kappa_{p}F_{2}(q^{2}).
  \label{f3}
\end{equation}
In the
TL region, the center-of-mass (c.m.)~system is equivalent to the Breit frame since the helicities of
baryons are opposite for the spinors aligned in $G_{M}$ and are the same for the spinors aligned in
$G_{E}$.

In the SL region, FFs have been extracted by the Rosenbluth separation method~\cite{Rosenbluth}, as
well as, more recently, by the recoil proton polarization transfer method~\cite{Polarization}. The
latter has been applied to obtain the $\mu_{p}G_{E}/G_{M}$ ratio.  Results from the GEp-II
experiment at JLab's Hall A~\cite{jlab1,jlab2} for $\mu_{p}G_{E}/G_{M}$ show that this ratio
decreases rather quickly with increasing $Q^{2}$, where $Q^{2}=-q^{2}\geq0$, while results achieved
by the Rosenbluth method show an almost constant ratio~\cite{And}.  The discrepancy between the
Rosenbluth and the polarization transfer method may be resolved by including higher order
corrections like two-photon exchange. A small correction to the Rosenbluth separation could imply a
large correction for the extraction of $G_{E}$, since $G_{E}$ is the slope of the Rosenbluth plot.
However, the correction of including two-photon exchange is small and cannot significantly influence
the results of the polarization transfer experiment.

In the TL region, measurements have been performed in the direct production channel
$e^{+}e^{-}\rightarrow p\bar{p}$~\cite{DM1,DM2,FENICE,bes2,cleoform}, in the radiative return
channel $e^{+}e^{-}\rightarrow p\bar{p}(\gamma_{ISR})$~\cite{babar2,babar3} where $\gamma_{ISR}$
refers to a photon emitted by initial state radiation (ISR), and in
$\bar{p}p\rightarrow e^{+}e^{-}$~\cite{ps170, E760,E835} experiments.  In cases where the data sample is
too small to extract angular distributions and disentangle $|G_{E}|$ and $|G_{M}|$, the effective
proton FF $|G|$ can be calculated from the total cross section, assuming $|G_{E}|=|G_{M}|$.  This
assumption is valid at the $p\bar{p}$ mass threshold, if analyticity of the FFs holds, implying that at
threshold the angular distribution should be isotropic.  In the PS170 experiment at
LEAR~\cite{ps170}, the effective proton FF was obtained, as well as the $|G_{E}/G_{M}|$ ratio, from
$p\bar{p}$ threshold up to $\sqrt{s}=2.05$ GeV.  In the BaBar experiment at
PEP-II~\cite{babar2,babar3}, the cross section was measured using the ISR method from the $p\bar{p}$
production threshold up to $\sqrt{s}=6.5$ GeV. The $|G_{E}/G_{M}|$ ratio was measured from threshold
up to $\sqrt{s}=3.0$ GeV and the result shows an inconsistency with respect to the PS170 results.

The presence of vector resonances, like $\rho$, $\omega$ and $\phi$ in the \emph{unphysical region},
below the $p\bar{p}$ threshold, can influence the functional form of the FFs in the physical region.
Hence the FFs, in particular the ratio $|G_{E}/G_{M}|$, in the TL region cannot be simply
extrapolated from the SL ones.  Until now it has been assumed that all FFs respect analyticity,
which should allow to calculate their behavior in the unphysical region thanks to dispersion
relations~\cite{dispersion} using the available data in both the TL and SL regions. In the SL
region, the ratio $\mu_{p}G_{E}/G_{M}$ has been measured at 16 $Q^{2}$ values in (0.5, 8.5)
GeV$^{2}$ with the best precision to 1.7$\%$~\cite{jlab1,jlab2}, while the present precision of
$|G_{E}/G_{M}|$ in the TL region exceeds $10\%$ by far. Therefore, it is necessary to improve
the measurement of $|G_{E}/G_{M}|$ ratio in the TL region.

The experimental determinations of proton FFs are important input for various QCD-based theoretical models.  There
are plenty of theoretical approaches applied to explain TL FFs: Chiral Perturbation
Theory~\cite{chiral}, Lattice QCD~\cite{lattice}~\cite{lattice2}, Vector Meson Dominance
(VMD)~\cite{VMD}, the Relativistic Constituent Quark Model (CQM)~\cite{CQM}, and, at high energies,
perturbative QCD predictions~\cite{pQCD}.

In this paper, we present an investigation of the process $e^{+}e^{-}\rightarrow p\bar{p}$ based on
data samples collected with the Beijing Spectrometer III (BESIII)~\cite{BESIII} at the Beijing
Electron Positron Collider II (BEPCII) at 12 c.m.~energies ($\sqrt{s}$).  The Born cross section at
these energy points are measured and the corresponding effective FFs are determined.  The ratio of
electric to magnetic FFs, $|G_{E}/G_{M}|$, and $|G_M|$ are measured at those c.m.~energies where the
statistics are large enough. The results are consistent
with those from BaBar in the same $q^2$ region.

\section{\boldmath The BESIII Experiment And Data Sets}
BEPCII is a double-ring $e^{+}e^{-}$ collider running at c.m.~energies between 2.0-4.6 GeV and
reached a peak luminosity of $0.85\times10^{33}$cm$^{-2}$s$^{-1}$ at a c.m.~energy of 3770 MeV.  The
cylindrical BESIII detector has an effective geometrical acceptance of $93\%$ of 4$\pi$ and is
divided into a barrel section and two endcaps.  It contains a small cell, helium-based (40$\%$ He,
60$\%$ C$_{3}$H$_{8}$) main drift chamber (MDC) which provides momentum measurement for charged
particles with a resolution of $0.5\%$ at a momentum of 1 GeV/c in a magnetic field of 1 Tesla.  The
energy loss measurement ($dE/dx$) provided by the MDC has a resolution better than $6\%$.  A
time-of-flight system (TOF) consisting of 5-cm-thick plastic scintillators can measure the flight
time of charged particles with a time resolution of 80 ps in the barrel and 110 ps in the end-caps.
An electromagnetic calorimeter (EMC) consisting of 6240 CsI (Tl) in a cylindrical structure and two
end-caps is used to measure the energies of photons and electrons.  The energy resolution of the EMC
is $2.5\%$ in the barrel and $5.0\%$ in the end-caps for photon/electron of 1 GeV energy.  The
position resolution of the EMC is 6 mm in the barrel and 9 mm in the end caps. A muon system (MUC)
consisting of about 1000~m$^{2}$ of Resistive Plate Chambers (RPC) is used to identify muons and provides
a spatial resolution better than 2~cm.

Monte Carlo (MC) simulated signal and background samples are used to optimize the event selection
criteria, estimate the background contamination and evaluate the selection efficiencies.  The MC
samples are generated using a {\sc Geant4}-based~\cite{geant4} simulation software package {\sc
  BESIII Object Oriented Simulation Tool (BOOST})~\cite{Deng}, which includes the description of
geometry and material, the detector response and the digitization model, as well as a database of
the detector running conditions and performances.  In this analysis, the generator software package
{\sc Conexc}~\cite{ping} is used to simulate the signal MC samples $e^{+}e^{-}\rightarrow p\bar{p}$,
and calculate the corresponding correction factors for higher order process with one radiative
photon in the final states.  Another generator {\sc Phokhara}~\cite{phokhara} serves as a cross
check of the radiative correction factors.  At each c.m.~energy, a large signal MC sample
with more than 10 times of the produced events in data for the process
$e^{+}e^{-}\rightarrow p\bar{p}$, contributing $0.15\%$ statistical uncertainty on
the detection efficiency, is generated.  Simulated samples of the QED background processes
$e^{+}e^{-}\rightarrow l^{+}l^{-}$ (l = e, $\mu$) and $e^{+}e^{-}\rightarrow \gamma\gamma$ are
generated with the generator {\sc Babayaga}~\cite{babayaga}.  The other background MC samples for
the processes with the hadronic final states $e^{+}e^{-}\rightarrow h^{+}h^{-}$ ($h = \pi$, $K$),
$e^{+}e^{-}\rightarrow p\bar{p}\pi^{0}$, $e^{+}e^{-}\rightarrow p\bar{p}\pi^{0}\pi^{0}$ and
$e^{+}e^{-}\rightarrow \Lambda\bar{\Lambda}$ are generated with uniform phase space distributions.
The background samples are generated with equivalent luminosities at least as large as the
data samples.

\section{\boldmath Analysis Strategy}
\subsection{\boldmath Event selection}
Charged tracks are reconstructed with the hit information from the MDC.
A good charged track must be within the MDC coverage, $|\cos\theta|<0.93$,
and is required to pass within 1 cm of the $e^{+}e^{-}$ interaction point (IP) in the plane
perpendicular to the beam and within $\pm10$ cm in the direction along the beam.
The combined information of $dE/dx$ and TOF is used to calculate particle identification (PID)
probabilities for the pion, kaon and proton hypothesis, respectively,
and the particle type with the highest probability is assigned to the track.
In this analysis, exactly two good charged tracks, one proton and one antiproton, are required.
To suppress Bhabha background events, the ratio $E/p$ of each proton candidate is required to be
smaller than 0.5, where $E$ and $p$ are the energy deposited in the EMC and the momentum measured
in the MDC, respectively.
The cosmic ray background is rejected by requiring $|T_\text{trk1}-T_\text{trk2}|<$4 ns, where $T_\text{trk1}$ and
$T_\text{trk2}$ are the measured time of flight in the TOF detector for the two tracks.
For the samples with c.m. energy $\sqrt{s}>2400.0$ MeV, the proton is further required to satisfy $\cos\theta<0.8$ to suppress Bhabha background.

\begin{figure*}[htbp]
\begin{center}
\begin{overpic}[width=3.in]{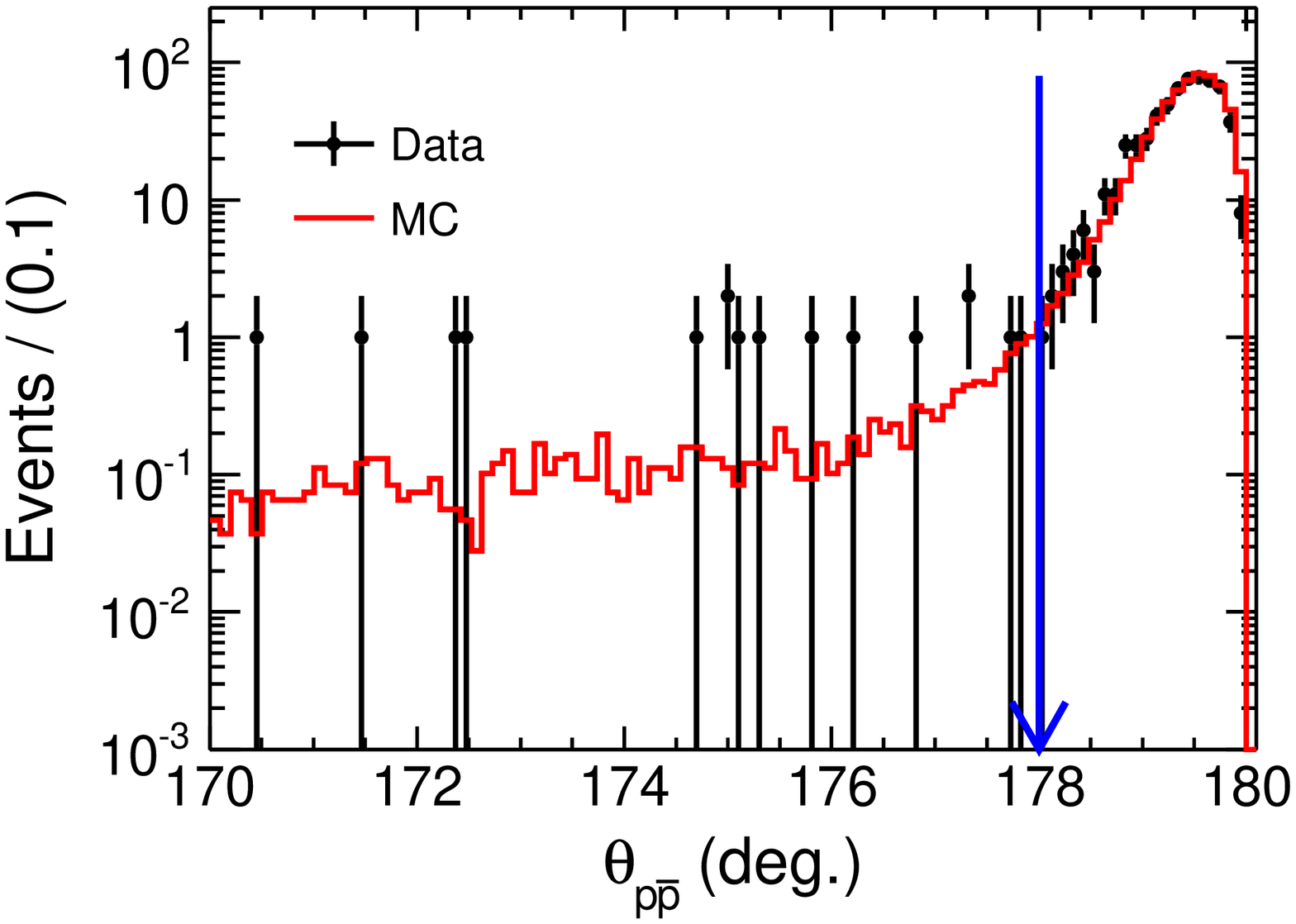}
\put(70,60){(a)}
\end{overpic}
\begin{overpic}[width=3.in]{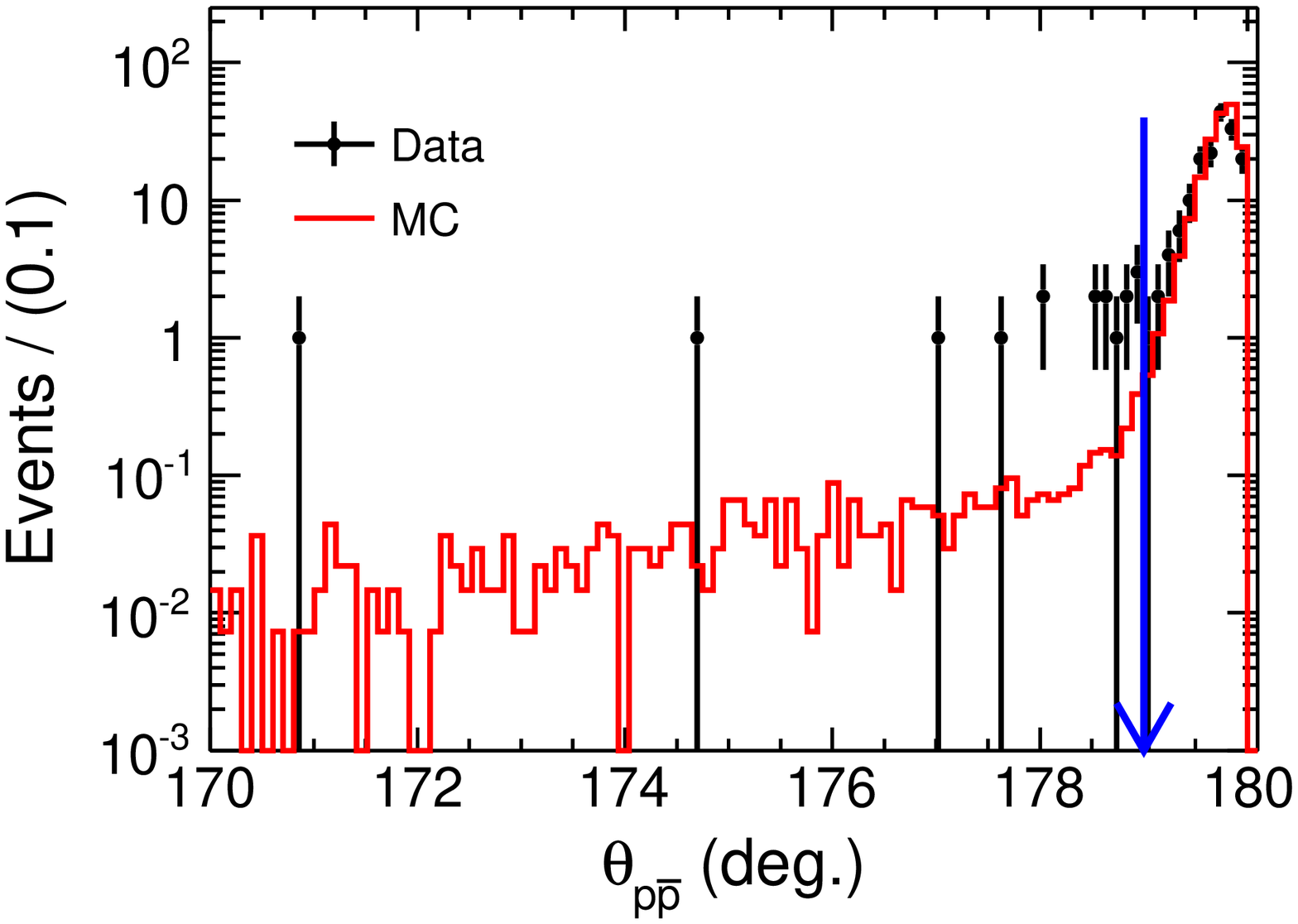}
\put(70,60){(b)}
\end{overpic}
\vskip -0.3cm
\parbox[1cm]{16cm} {
\caption{Opening angle distributions between proton and antiproton at the c.m.~energies of (a) 2232.4 MeV, and (b) 3080.0 MeV.
The dots with error bars are data, the histograms represent the distributions of
signal MC samples. The arrows show the selection applied.}
\label{angle}}
\end{center}
\end{figure*}

\begin{figure*}[htbp]
\begin{center}
\begin{overpic}[width=3.in]{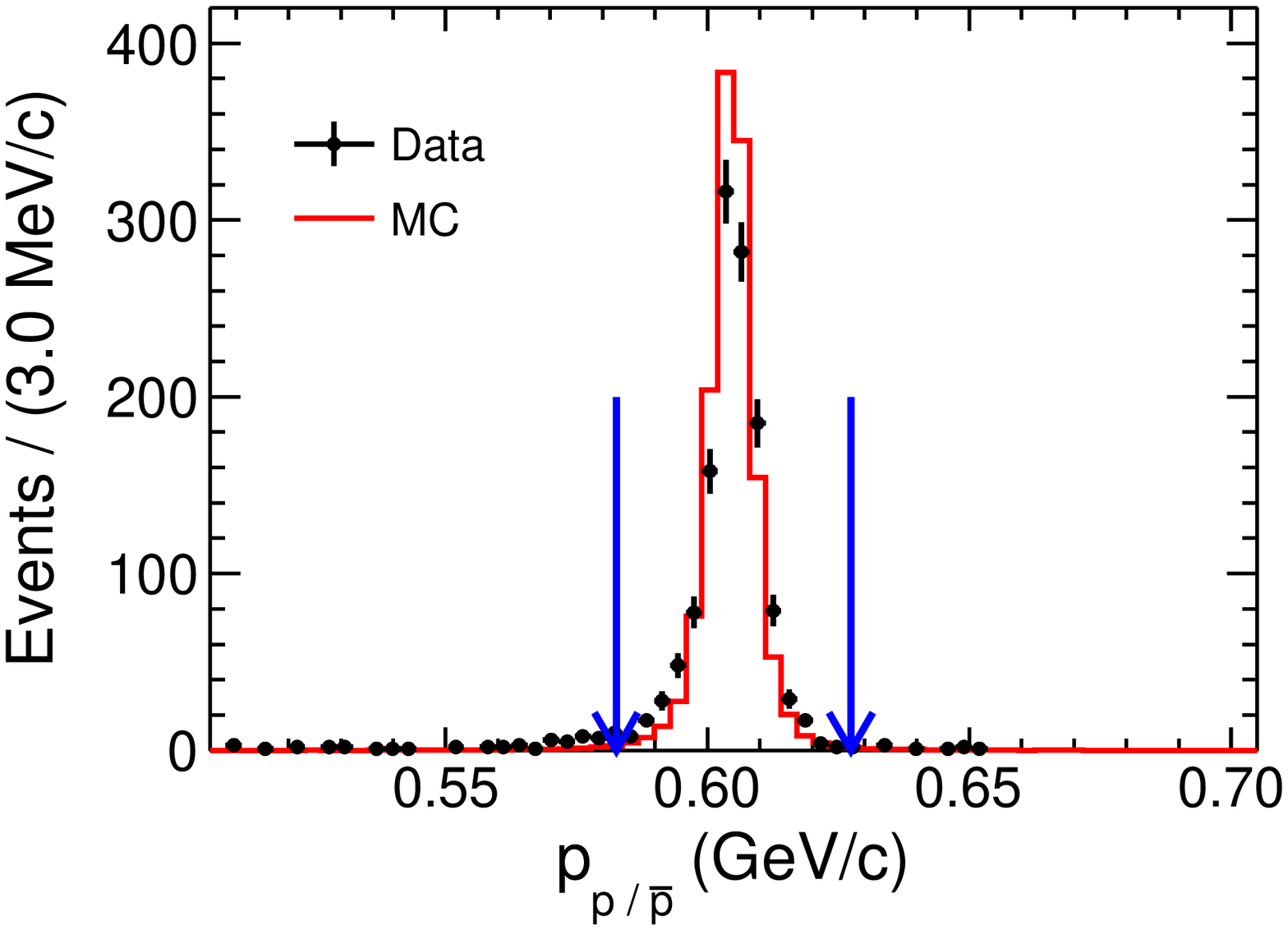}
\put(70,60){(a)}
\end{overpic}
\begin{overpic}[width=3.in]{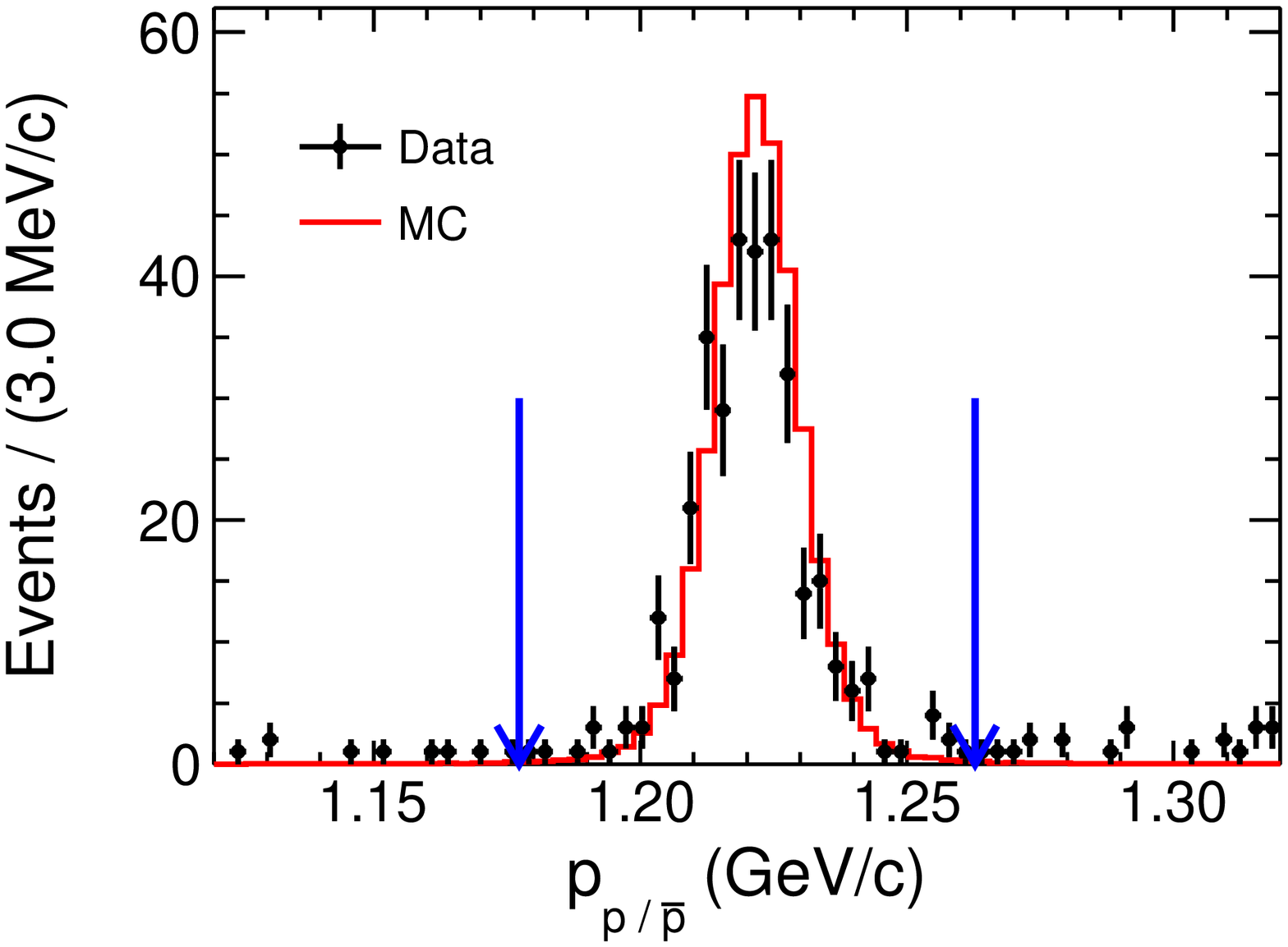}
\put(70,60){(b)}
\end{overpic}
\vskip -0.3cm
\parbox[1cm]{16cm} {
\caption{Momentum distribution of the proton or antiproton at the c.m.~energies (a) 2232.4 MeV,
and (b) 3080.0 MeV, two entries per event.
The dots with error bars are data, the histograms represent the distributions of signal MC samples.
The arrows show the momentum window requirements. }
\label{mom}}
\end{center}
\end{figure*}

After applying the above selection criteria, the distributions of the opening angle between
proton and antiproton, $\theta_{p\bar{p}}$, at c.m.~energies $\sqrt{s}=$ 2232.4 and 3080.0 MeV
are shown in Fig.~\ref{angle}.
Good agreement between data and MC samples is observed, and a better resolution is achieved
with increasing c.m.~energy due to the smaller effects on the small angle multiple scattering.
A c.m.~energy dependent requirement, $i.e.$, $\theta_{p\bar{p}}>178^\circ$ at $\sqrt{s}\leq2400.0$
MeV, and $\theta_{p\bar{p}}>179^\circ$ at $\sqrt{s}>2400.0$ MeV, is further applied.
Figure~\ref{mom} shows the distribution of the momentum of proton or antiproton
at c.m.~energies $\sqrt{s}=2232.4$ and 3080.0 MeV.
A momentum window of 5 times the momentum resolution, $|p_\text{mea}-p_\text{exp}|<5\sigma_{p}$, is applied to extract
the signals, where $p_\text{mea}$ and $p_\text{exp}$ are the measured and expected momentum of the proton or antiproton
in the c.m.~system, respectively, and $\sigma_{p}$ is the corresponding resolution.

\subsection{\boldmath Background study}

The potential background contamination can be classified into two categories, the beam associated background and the physical background.

The beam associated background includes interactions between the beam and the
beam pipe, beam and residual gas, and the Touschek effect~\cite{touschek}.
Dedicated data samples with separated beams were collected with the BESIII detector at $\sqrt{s}=$ 2400.0 and 3400.0 MeV;
these are used to study the beam associated background.
Since the two beams do not interact with each other, all of the observed events are beam associated
background, and can be used to evaluate the beam associated background at different c.m.~energies by
normalizing the data-taking time and efficiencies.
No events from the separated beam data samples survive the signal selection criteria.
Considering that the normalization factor is less than 5 for most of energy points (other
than 3.08 and 3.65~GeV), the beam associated background at all c.m.~energy points is negligible.

The physical background may come from the  $e^{+}e^{-}$ annihilation  processes with two-body final states, {\it e.g.}
Bhabha or di-muon events, where leptons are misidentified as protons or antiprotons, or processes with
multi-body final states including $p\bar{p}$, {\it e.g.} $e^+e^-\to p\bar{p} \pi^0 (\pi^0)$.
The contamination from physical background is evaluated
by MC samples, and are listed in Table~\ref{bkg} for $\sqrt{s}=2232.4$ and $3080.0$~MeV, respectively.

The number of the surviving background events after normalization, $N^\text{MC}_\text{nor}$, is very small at the low c.m.~energies and can
therefore be safely neglected.   However, at higher c.m.~energies ($\sqrt{s}\geq3.40$~GeV), due to the rapid decrease of the cross section of
$e^{+}e^{-}\rightarrow p\bar{p}$, the background level which is mainly from Bhabha events is higher, and $N^\text{MC}_\text{nor}$ needs to
be corrected for.

\begin{figure*}[htbp]
\begin{center}
\begin{overpic}[width=0.32 \textwidth,angle=0]{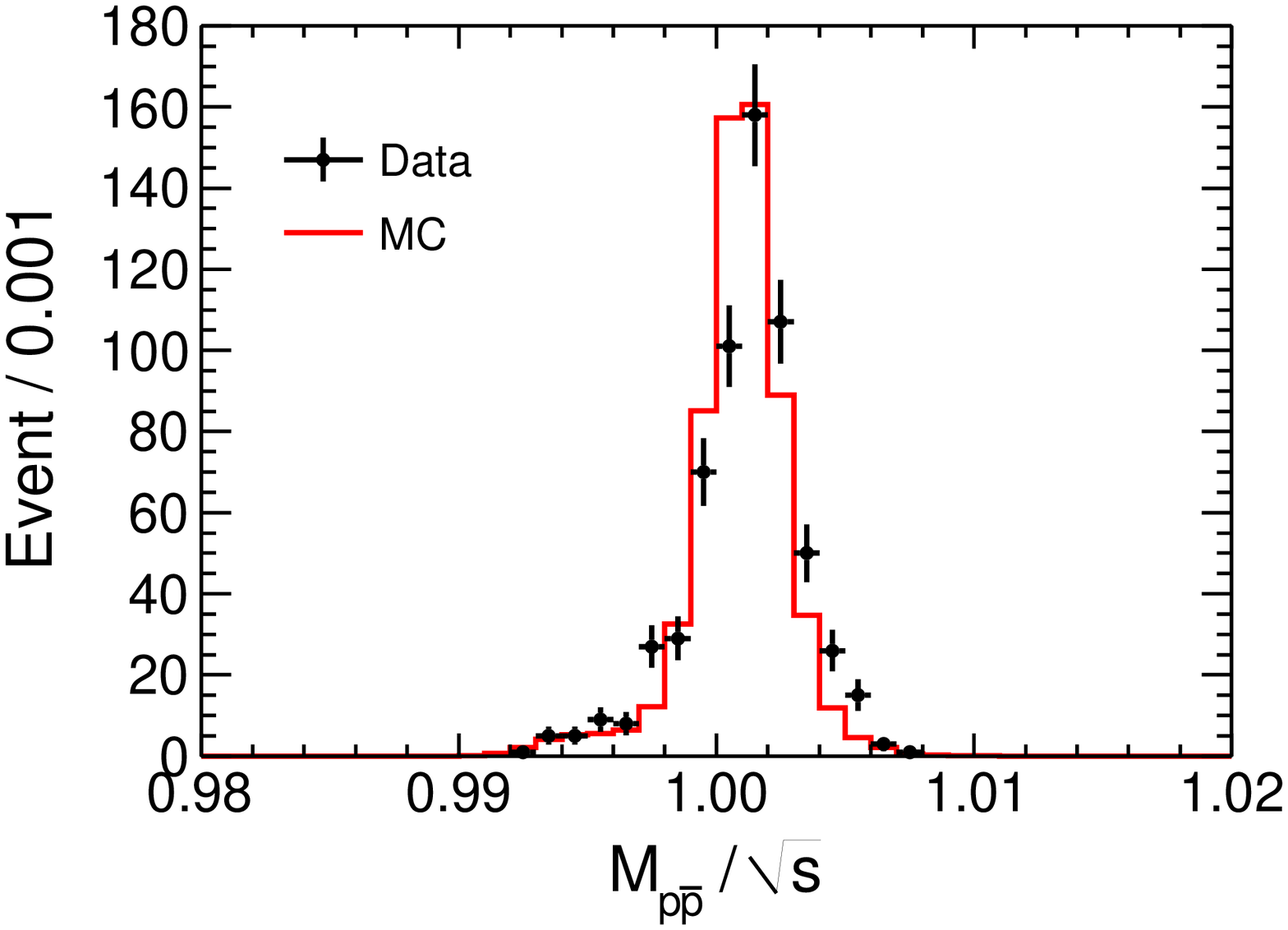}
\put(70,60){\footnotesize(a)}
\end{overpic}
\begin{overpic}[width=0.32 \textwidth,angle=0]{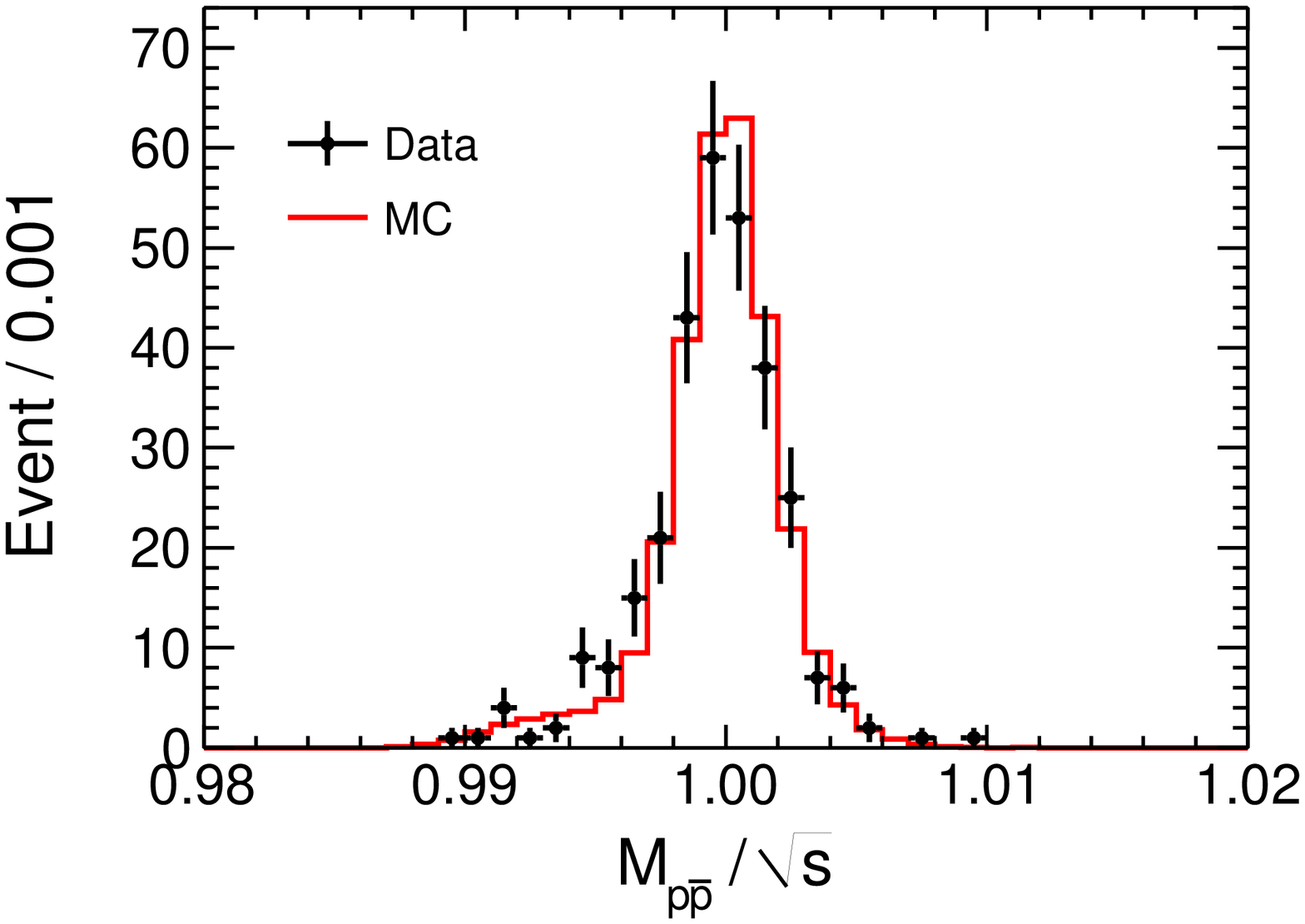}
\put(70,60){\footnotesize(b)}
\end{overpic}
\begin{overpic}[width=0.32 \textwidth,angle=0]{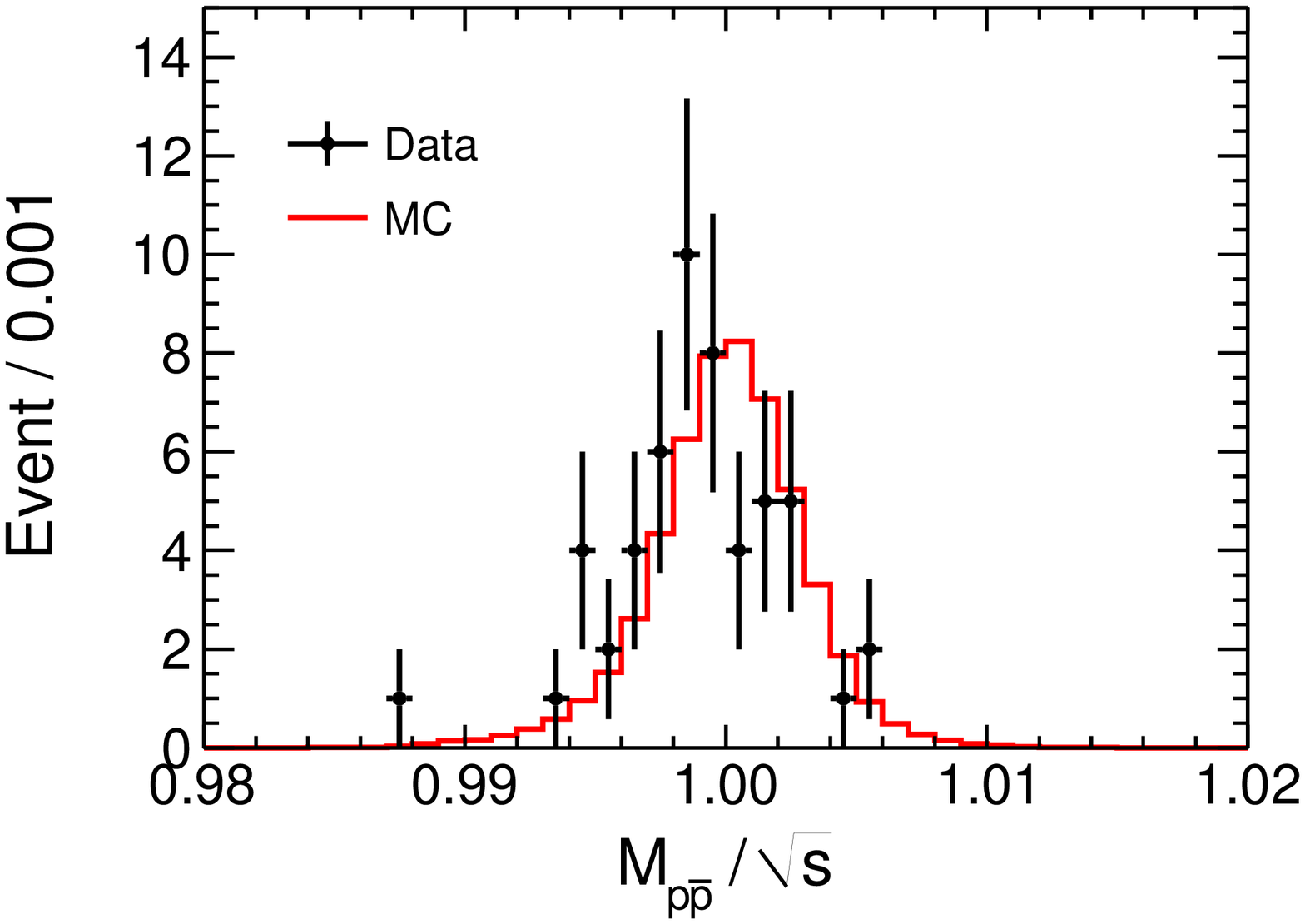}
\put(70,60){\footnotesize(c)}
\end{overpic}\\
\begin{overpic}[width=0.32 \textwidth,angle=0]{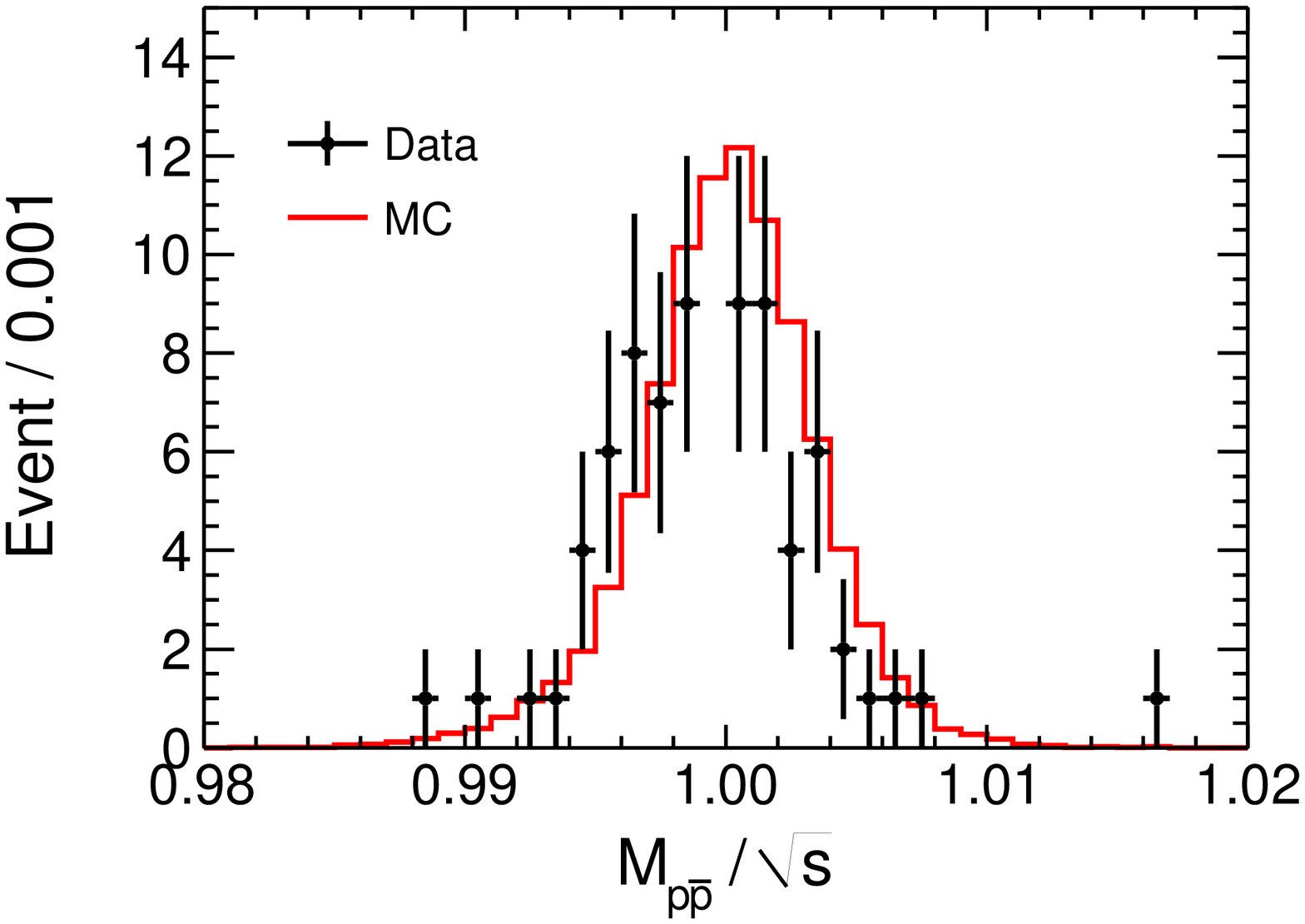}
\put(70,60){\footnotesize(d)}
\end{overpic}
\begin{overpic}[width=0.32 \textwidth,angle=0]{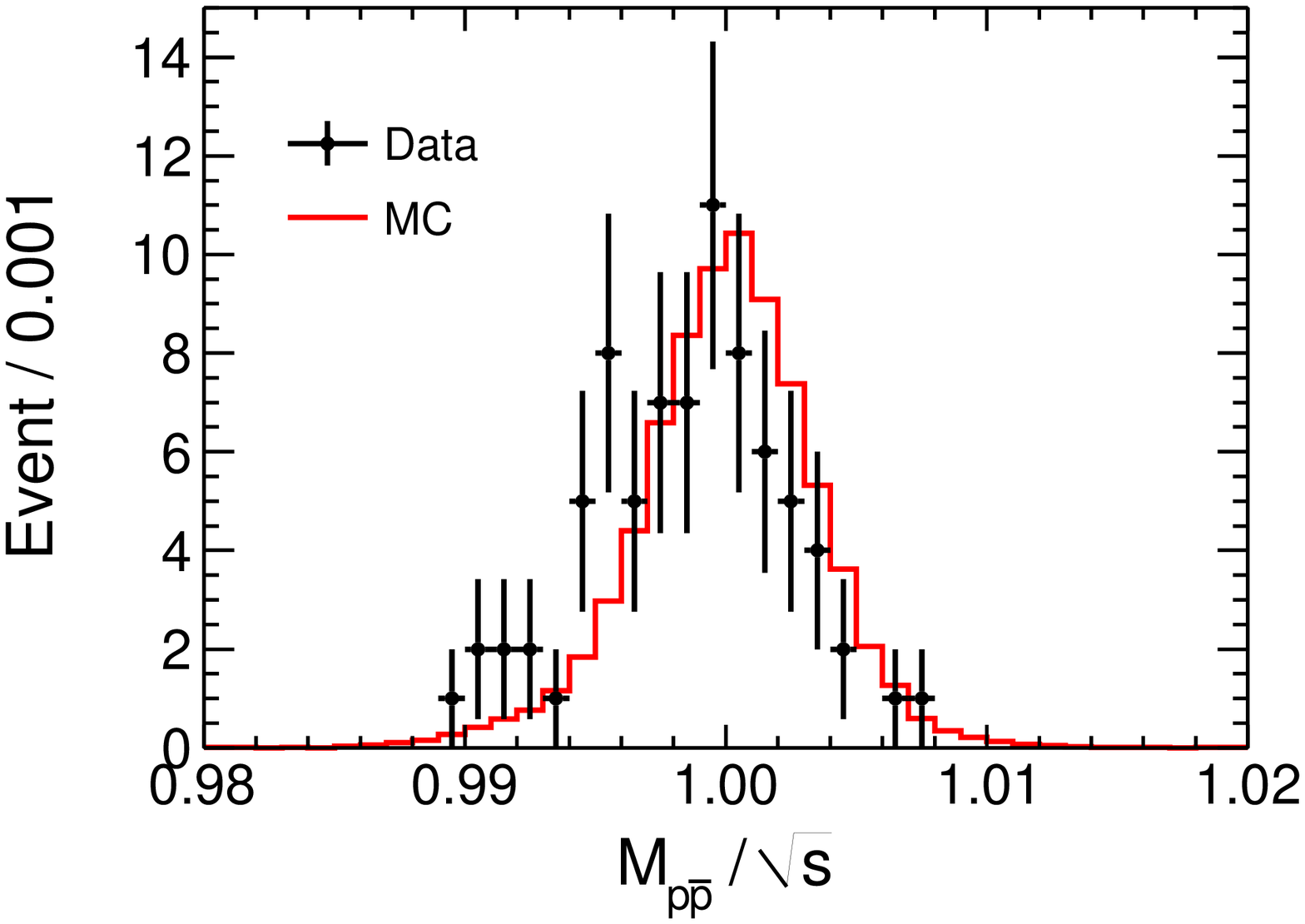}
\put(70,60){\footnotesize(e)}
\end{overpic}
\begin{overpic}[width=0.32 \textwidth,angle=0]{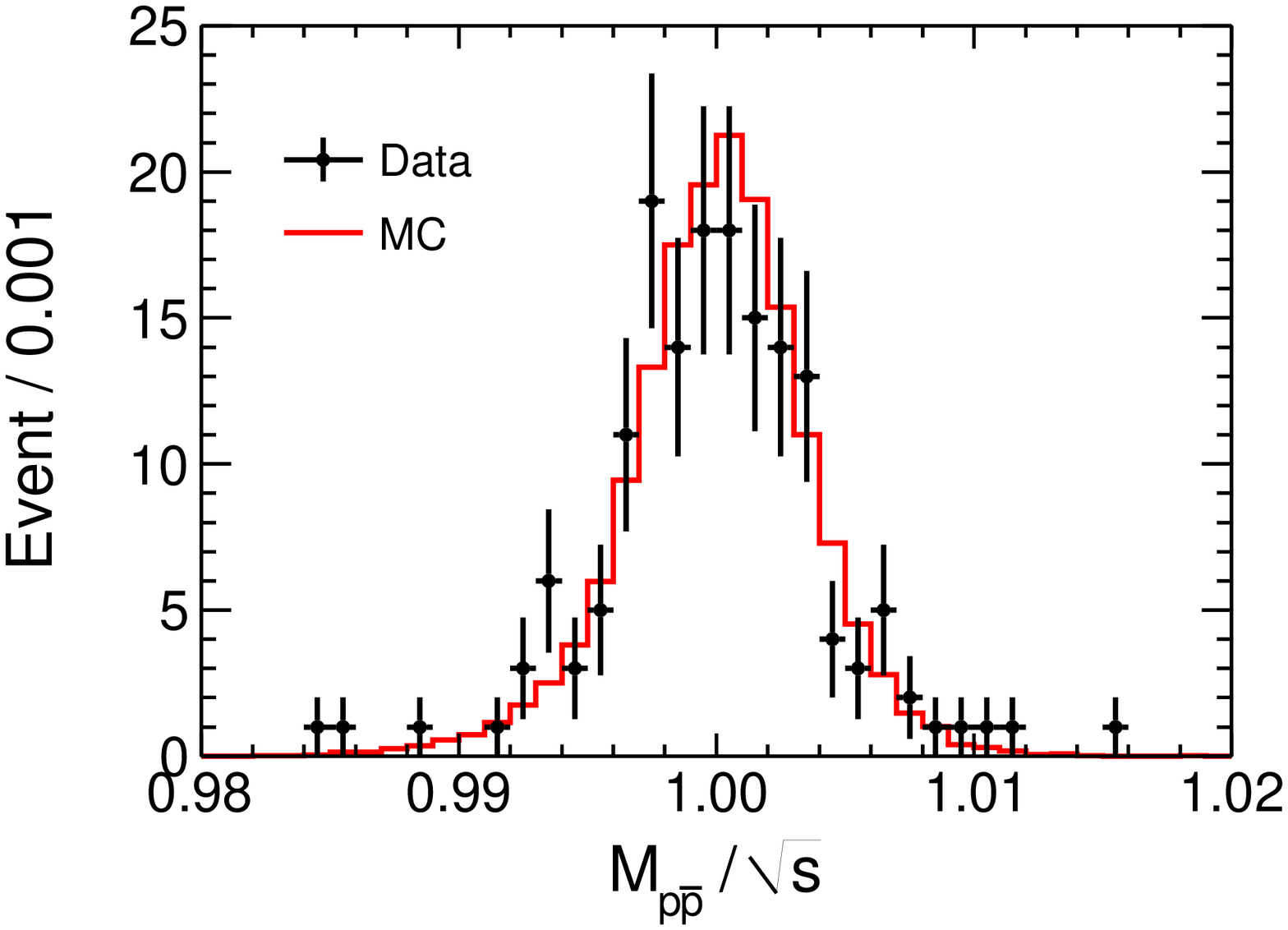}
\put(70,60){\footnotesize(f)}
\end{overpic}\\
\begin{overpic}[width=0.32 \textwidth,angle=0]{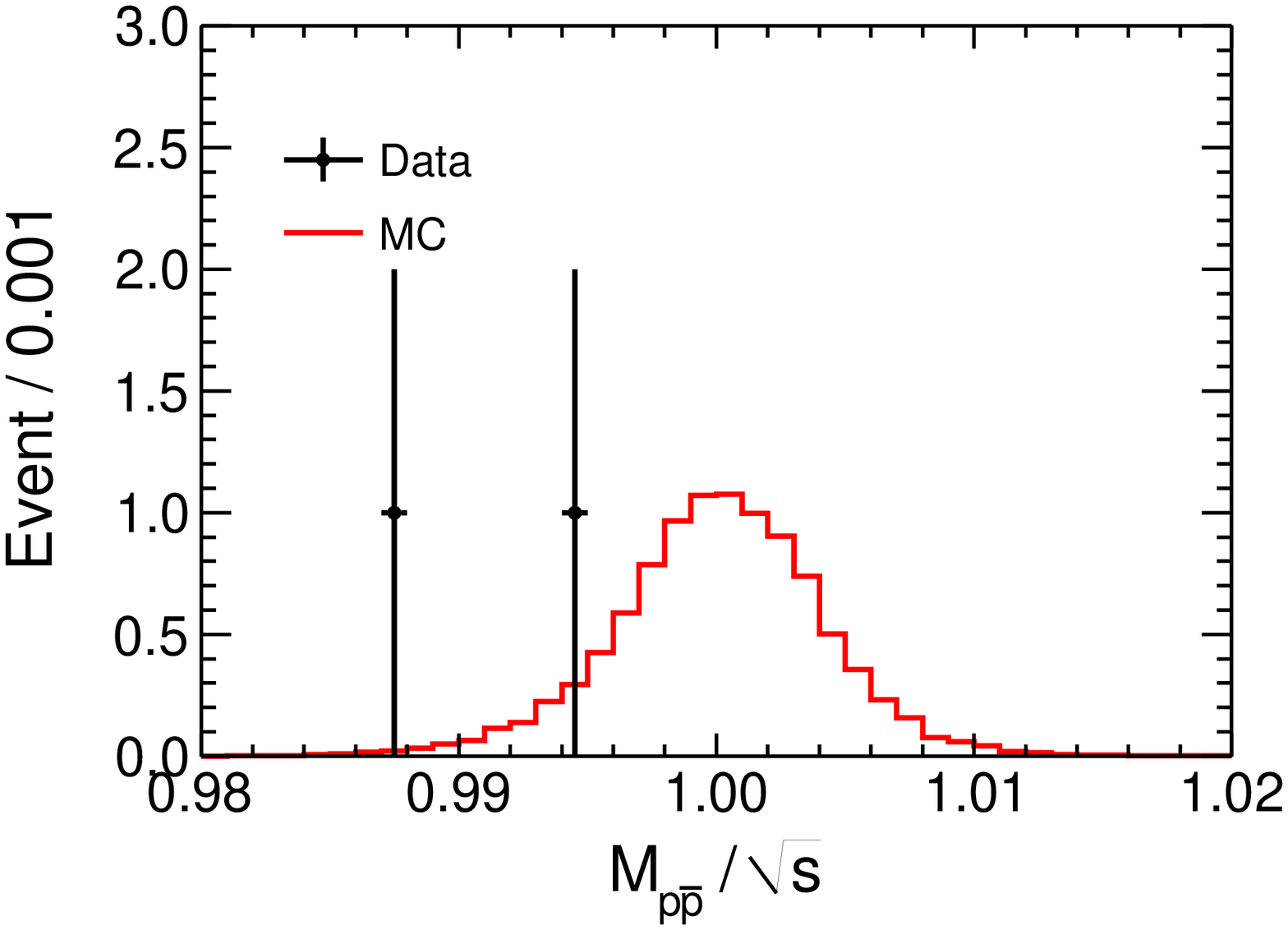}
\put(70,60){\footnotesize(g)}
\end{overpic}
\begin{overpic}[width=0.32 \textwidth,angle=0]{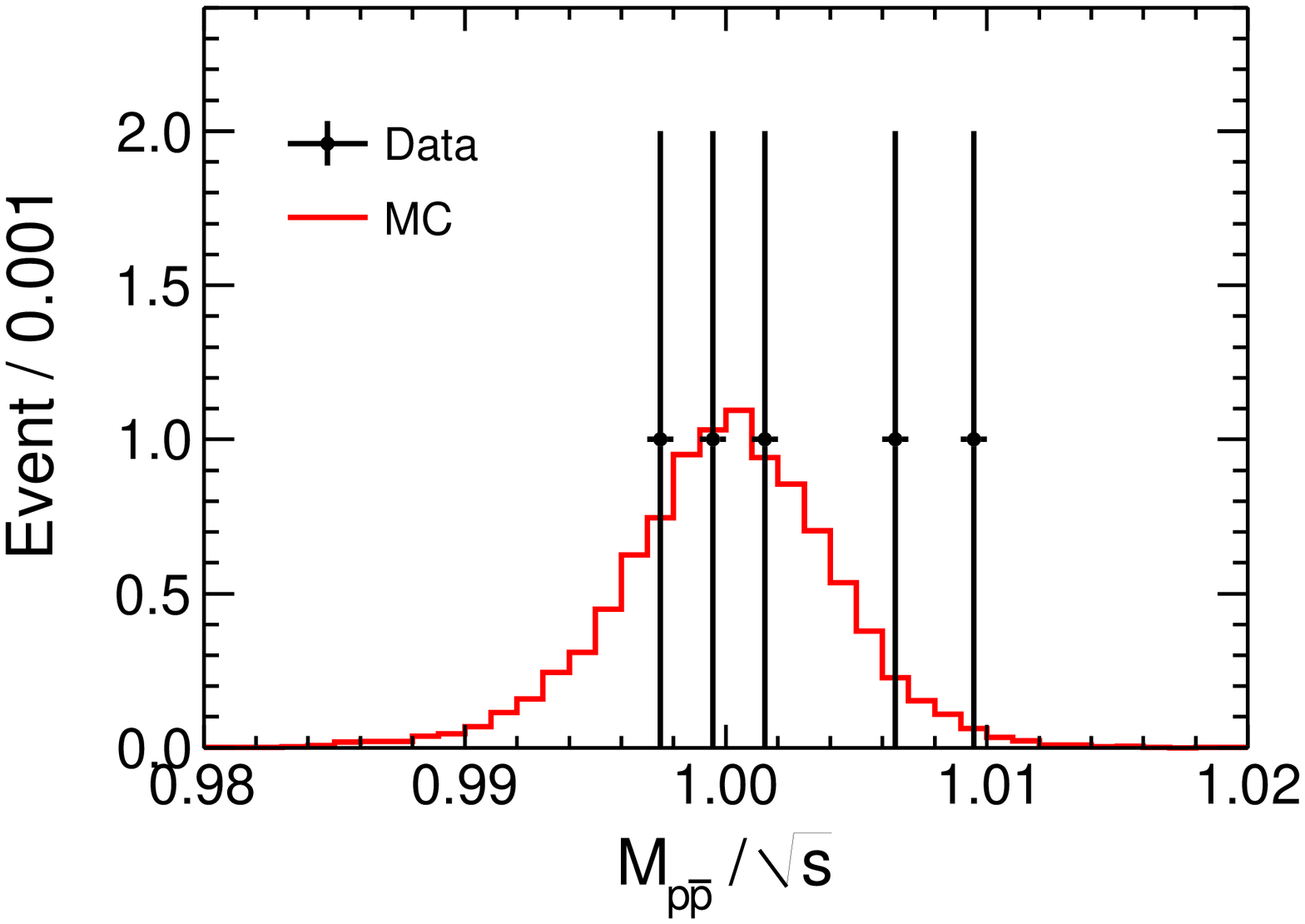}
\put(70,60){\footnotesize(h)}
\end{overpic}
\begin{overpic}[width=0.32 \textwidth,angle=0]{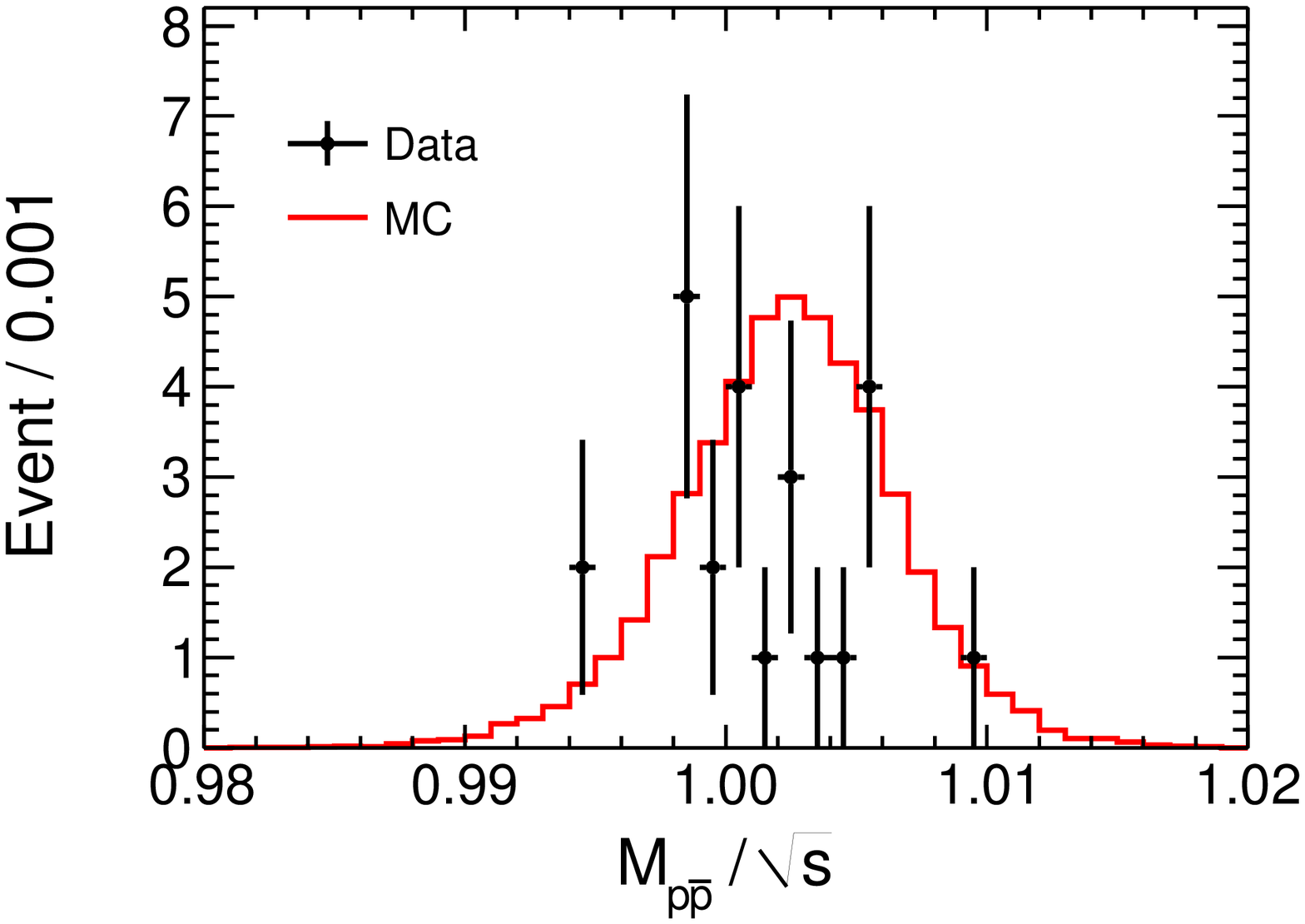}
\put(70,60){\footnotesize(i)}
\end{overpic}\\
\begin{overpic}[width=0.32 \textwidth,angle=0]{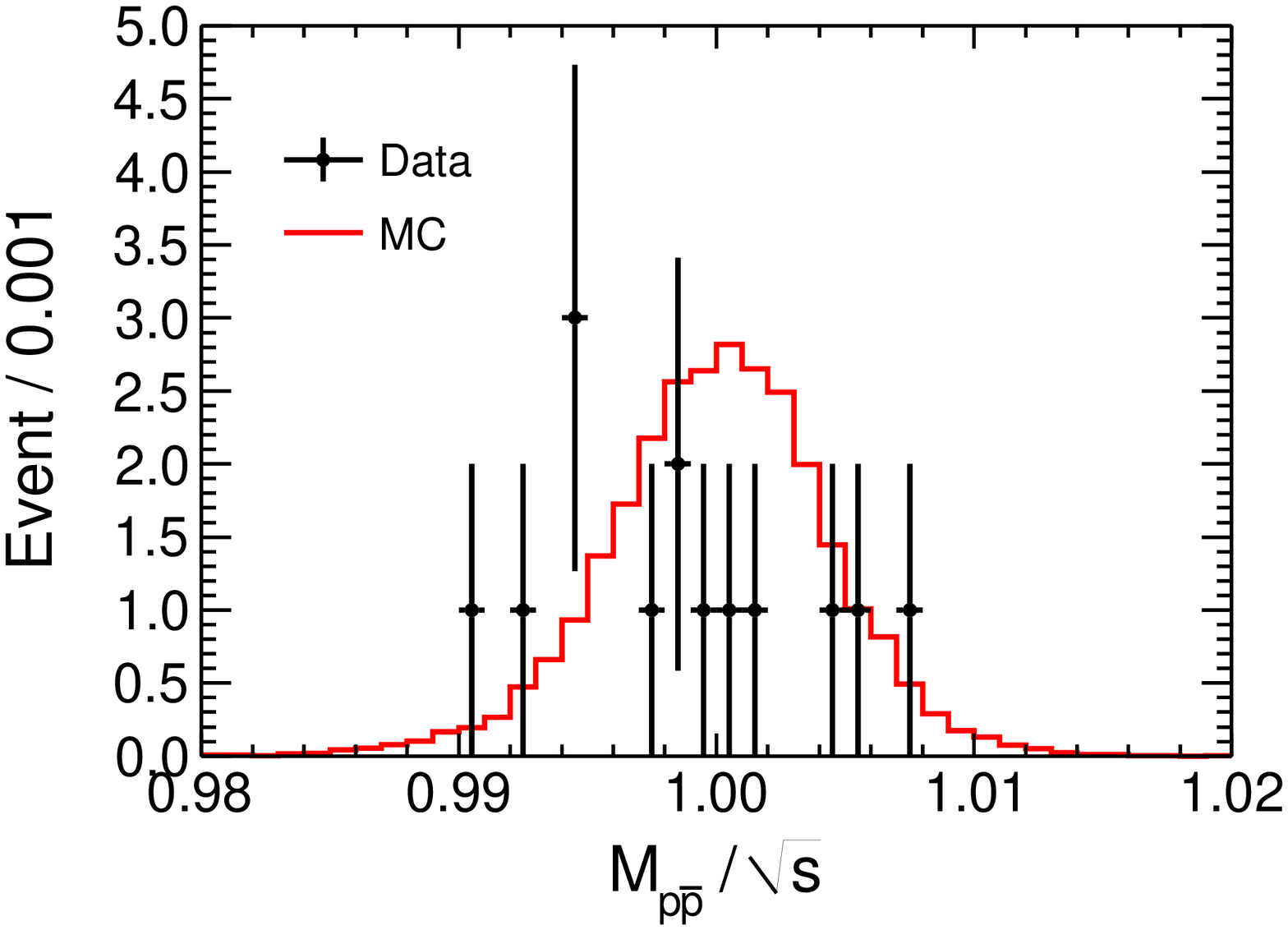}
\put(70,60){\footnotesize(j)}
\end{overpic}
\begin{overpic}[width=0.32 \textwidth,angle=0]{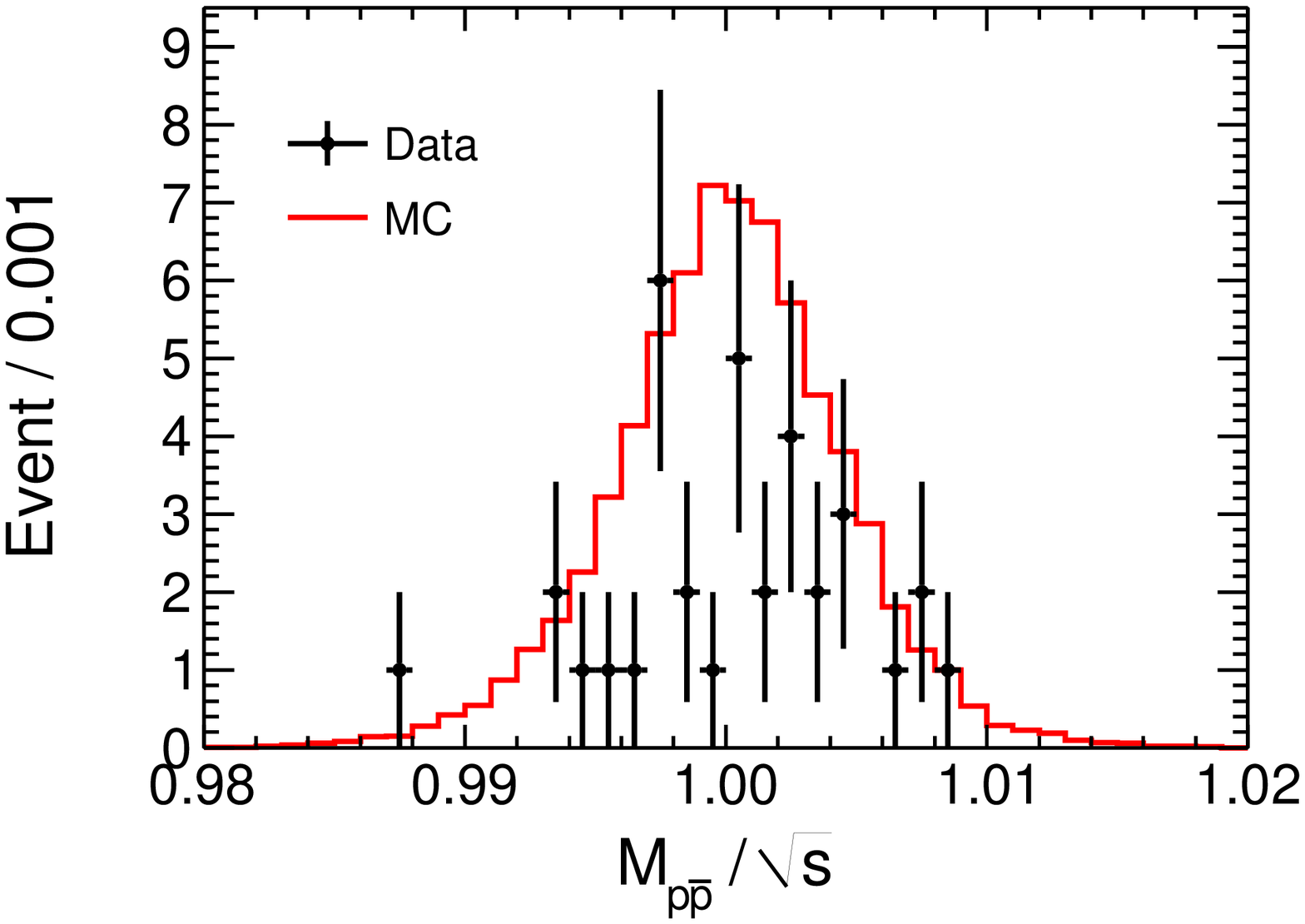}
\put(70,60){\footnotesize(k)}
\end{overpic}
\begin{overpic}[width=0.32 \textwidth,angle=0]{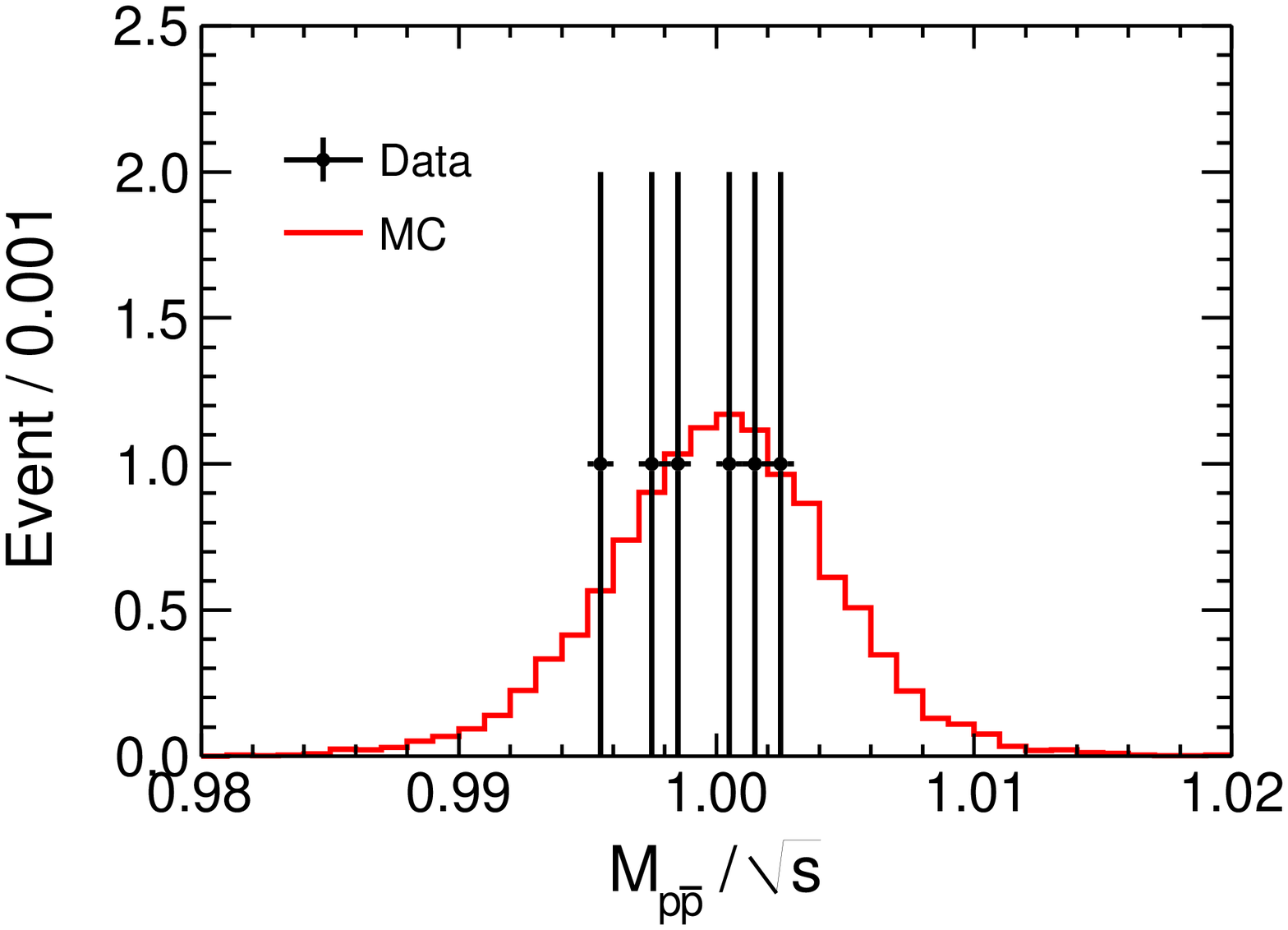}
\put(70,60){\footnotesize(l)}
\end{overpic}
 \caption{Comparison of $M_{p\bar{p}}/\sqrt{s}$ distributions at different c.m.~energies for data (dots with error bars)
 and MC (histograms):
 (a) 2232.4, (b) 2400.0, (c) 2800.0, (d) 3050.0, (e) 3060.0, (f) 3080.0,
 (g) 3400.0, (h) 3500.0, (i) 3550.7, (j) 3600.2, (k) 3650.0, (l) 3671.0 MeV.
 The sample (i) is a combination of three data sub-samples with very close c.m.~energies, $\sqrt{s}=$ 3542.4, 3553.8, 3561.1 MeV, and the
 value of 3550.7~MeV is the average c.m.~energy weighted with their luminosity values.
 }\label{mpp}
\end{center}
\end{figure*}

\begin{table*}[htbp]
\caption{Physical background processes estimated from the MC samples at $\sqrt{s}=2232.4$ and $3080.0$ MeV. $N^\text{MC}_\text{gen}$ is the
  number of generated MC events, $N^\text{MC}_\text{sur}$ is the number of events remaining after the selection criteria,
  $\sigma$ is the production cross section in the $e^{+}e^{-}$ annihilation process,
  which is obtained using the Babayaga generator for Bhabha, di-muon, and di-photon processes,
  and from the previous experimental results for others processes~\cite{xsection1,xsection2}.
  $N^\text{MC}_\text{uplimit}$ and $N^\text{MC}_\text{nor}$ are the estimated upper limit at the 90$\%$ confidence level (C.L.)
  and the normalized number of background events.
}
\footnotesize
\begin{center}
\begin{tabular}{c|ccccc|ccccc}
\hline
\hline
& \multicolumn{5}{|c} {{\small  $\sqrt{s} = 2232.4$ MeV ($2.63$~pb$^{-1}$)} } & \multicolumn{5}{|c} {{\small  $\sqrt{s} = 3080.0$ MeV ($30.73$~pb$^{-1}$)} } \\
\hline
\hline
Bkg. & $N^\text{MC}_\text{gen}$ ($\times10^{6}$) & $N^\text{MC}_\text{sur}$ & $\sigma$ (nb) & $N^\text{MC}_\text{uplimit}$
 & $N^\text{MC}_\text{nor}$ & $N^\text{MC}_\text{gen}$ ($\times10^{6}$) & $N^\text{MC}_\text{sur}$ & $\sigma$ (nb)
 & $N^\text{MC}_\text{uplimit}$ & $N^\text{MC}_\text{nor}$\\ \hline
$e^{+}e^{-}$ &     9.6 & 0 & 1435.01 & $<0.96$ & 0 & 39.9 & 1 & 756.86 & $<2.54$ & 1\\
$\mu^{+}\mu^{-}$ & 0.7 & 0 &  17.41  & $<0.16$ & 0 & 1.5  & 0 &8.45  & $<0.42$  & 0\\
$\gamma\gamma$ &   1.9 & 0 &  70.44  & $<0.24$ & 0 & 4.5  & 0 &37.05 & $<0.62$ & 0\\
$\pi^{+}\pi^{-}$ & 0.1 & 0 &  0.17  & $<0.01$ & 0 & 0.1  & 0 &$<0.11$ & $<0.02$ & 0\\
$K^{+}K^{-}$ &     0.1 & 0 &  0.14  & $<0.008$ & 0 & 0.1 & 0 &0.093 & $<0.02$ & 0\\
$p\bar{p}\pi^{0}$ &0.1 & 0 &  $<0.1$ & $<0.006$ & 0 & 0.1 & 0 &$<0.1$ & $<0.07$ & 0\\
$p\bar{p}\pi^{0}\pi^{0}$ & 0.1 & 0 & $<0.1$ & $<0.006$ & 0 & 0.1 &0 & $<0.1$ & $<0.07$ & 0\\
$\Lambda\overline{\Lambda}$ & 0.1 & 0 & $<0.4$ & $<0.02$ & 0 & 0.1 &0 & 0.002 & $<0.001$ & 0\\
\hline
\hline
\end{tabular}
    \label{bkg}
\end{center}
\end{table*}
The ratio of $p\bar{p}$ invariant mass and the c.m.~energy,
$M_{p \bar{p}}/\sqrt{s}$,
from data and MC has been compared and is shown in
Fig.~\ref{mpp} at different c.m.~energies.
The integral luminosity of the data set at each c.m.~energy is listed in Table~\ref{result1}.
There is good agreement between data and MC simulations.
The signal yields are extracted by counting the number of events and are listed in Table~\ref{result1},
where the quoted uncertainties are statistical only.
The data sample at $3550.7$~MeV is a combination of three data sub-samples with very close c.m.~energies, $\sqrt{s}=$ 3542.4,
3553.8, 3561.1 MeV, and the value of 3550.7~MeV is the average c.m.~energy weighted with their luminosity values.

\subsection{\boldmath { Extraction of the Born cross section of $e^{+}e^{-}\rightarrow p\bar{p}$ and the effective FF}}

The differential Born cross section of $e^{+}e^{-}\rightarrow p\bar{p}$ can be written as a function of FFs, $|G_{E}|$ and $|G_{M}|$~\cite{formula},
\begin{eqnarray}
\begin{split}
 \frac{d\sigma_\text{Born}(s)}{d\Omega}  =  \frac{\alpha^{2}\beta C}{4s}[ & |G_{M}(s)|^{2}(1+\cos^{2}\theta_{p}) +\\
& \frac{4m_{p}^{2}}{s}|G_{E}(s)|^{2}\sin^{2}\theta_{p}],
  \label{eq1}
\end{split}
\end{eqnarray}
where $\alpha \approx \frac{1}{137}$ is the fine structure constant, $\beta=\sqrt{1-\frac{4m_{p}^{2}}{s}}$
is the velocity of the proton in the $e^+e^-$ c.m.~system, $C=\frac{\pi\alpha}{\beta}\frac{1}{1-\exp(-\pi\alpha/\beta)}$
is the Coulomb correction factor for a point-like proton, $s$ is the square of the c.m.~energy,
and $\theta_p$ is the polar angle of the proton in the $e^+e^-$ c.m.~system.
We assume that the proton is point-like above the $p\bar{p}$ production threshold, meaning that the Coulomb force acts only on the already
formed hadrons. At the energies we are considering here, the Coulomb correction factor can be safely
assumed to be 1.
Furthermore, under the assumption of the effective FF $|G|=|G_{E}|=|G_{M}|$ and by integrating over $\theta_p$,
it can be deduced:
 \begin{equation}
 |G|= \sqrt{\frac{\sigma_\text{Born}}{86.83\cdot\frac{\beta}{s}(1+\frac{2m_{p}^{2}}{s})}},
   \label{eq4}
 \end{equation}
where $\sigma_\text{Born}$ is in nb and $m_{p}$, $s$ in GeV.

\begin{figure*}[htbp]
\begin{center}
\begin{overpic}[width=3.in]{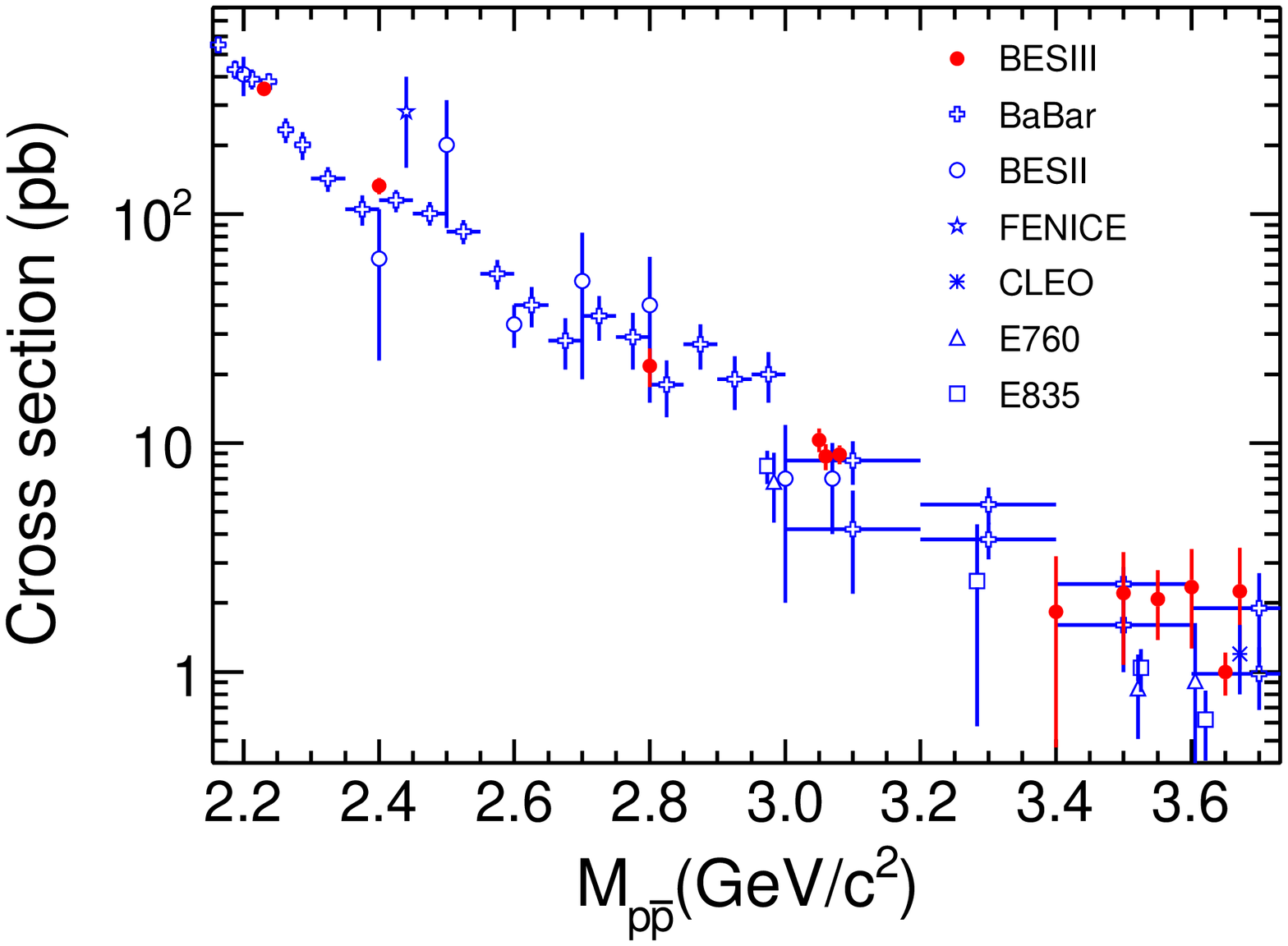}
\put(60,60){(a)}
\end{overpic}
\begin{overpic}[width=3.in]{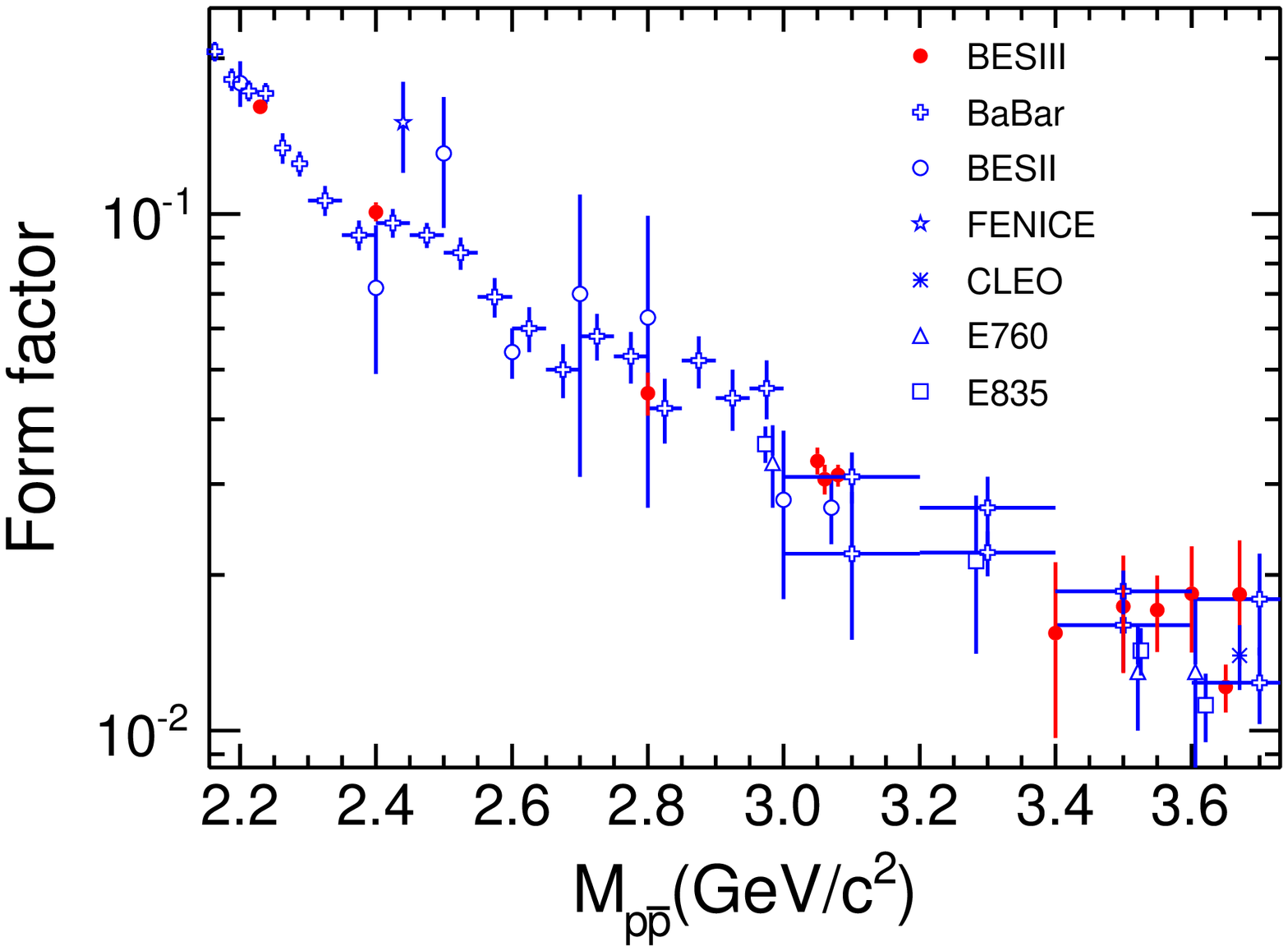}
\put(60,60){(b)}
\end{overpic}\\
 \caption{Comparison of (a) the Born cross section and (b) the effective FF $|G|$ between this measurement and previous experiments,
 shown on a logarithmic scale for invariant $p\bar{p}$
 masses from 2.20 to 3.70 GeV/$c^{2}$. }\label{compare1}
\end{center}
\end{figure*}

Experimentally, the Born cross section of $e^{+}e^{-}\rightarrow p\bar{p}$ is calculated by
\begin{equation}
  \sigma_\text{Born}=\frac{N_\text{obs}-N_\text{bkg}}{L\cdot\varepsilon\cdot(1+\delta)},
    \label{eq5}
\end{equation}
where $N_\text{obs}$ is the observed number of candidate events, extracted by counting the number of signal events,
$N_\text{bkg}$ is the expected number of background events estimated by MC simulations,
$L$ is the integrated luminosity estimated with large-angle Bhabha events,
$\varepsilon$ is the detection efficiency determined from a MC sample generated using the {\sc Conexc} generator~\cite{ping}, which includes
radiative corrections (which will be discussed in detail in next paragraph),
and $(1+\delta)$ is the radiative correction factor which has also been determined using the {\sc Conexc} generator.
The derived Born cross section $\sigma_\text{Born}$, the effective FF $|G|$, as well as the related variables used to calculate $\sigma_\text{Born}$
are shown in Table~\ref{result1} at different c.m.~energies.
In the table, the product value $\varepsilon'=\varepsilon\times(1+\delta)$ is presented
to account for the effective efficiency.
Comparisons of $\sigma_\text{Born}$ and $|G|$ to the previous experimental measurements are shown in Fig.~\ref{compare1}.
Compared to the BaBar results~\cite{babar2}, the precision of the Born cross section is improved by
30\% for data sets with $\sqrt{s}\leq3080.0$ MeV, and the corresponding precision of effective FF is improved, too.

From Eq.~\ref{eq1}, it is obvious that the detection efficiency depends on the ratio of the electric and magnetic FFs, $|G_E/G_M|$,
due to the different polar angle $\theta_p$ distributions. In this analysis, the detection efficiency is evaluated with the MC samples.
The ratio of $|G_E/G_M|$ is measured for data samples at c.m.~energies $\sqrt{s}$
= 2232.4 and 2400.0~MeV,  and for a combined data with sub-data samples at $\sqrt{s}$ = 3050.0,
3060.0, and 3080.0 MeV, which have close c.m.~energy. The corresponding measured $|G_E/G_M|$
ratios are used as the inputs for MC generation.
Details of the $|G_E/G_M|$ ratio measurement can be found in Sec.~\ref{secd}.
For other c.m.~energy points, where the $|G_E/G_M|$ ratios are not measured due to
limited statistics, the detection efficiencies are obtained by averaging the efficiencies with setting
$|G_E|=0$ and $|G_M|=0$, respectively.
The corresponding product values of detection efficiencies and the radiative correction factors at different c.m.~energies are listed in Table~\ref{result1}.
The interference of $p\bar{p}$ final states between $e^{+}e^{-}$ annihilation and $J/\psi$ decay in the lower tail is assumed to be negligible~\cite{bian}.

\begin{table*}[htbp]
\caption{Summary of the Born cross section $\sigma_\text{Born}$, the effective FF $|G|$, and the related variables used to calculate the Born cross sections at
the different c.m.~energies $\sqrt{s}$, where $N_\text{obs}$ is the number of candidate events, $N_\text{bkg}$
is the estimated background yield, $\varepsilon'=\varepsilon\times(1+\delta)$ is the product of detection efficiency $\varepsilon$ and the radiative
correction factor $(1+\delta)$, and $L$ is the integrated luminosity.
The first errors are statistical, and the second systematic.}
\footnotesize
\begin{center}
\begin{tabular}{c|cccccccc}
  \hline
  \hline
  $\sqrt{s}$ (MeV) & $N_\text{obs}$ &$N_\text{bkg}$ &  $\varepsilon'$  $(\%)$
  & $L$ (pb$^{-1}$)   & $\sigma_\text{Born}$ (pb) & $|G|$ $(\times10^{-2})$   \\
  \hline
   2232.4&  $614\pm25$ & 1 & 66.00 & 2.63 &    $353.0\pm14.3\pm15.5$ &  $16.10\pm0.32\pm0.35$\\
   2400.0 &  $297\pm17$& 1 & 65.79 &3.42  &  $132.7\pm7.7\pm8.1$ &  $10.07\pm0.29\pm0.31$\\
   2800.0 &  $53\pm7$ &1  &  65.08 &3.75 &  $21.3\pm3.0\pm2.8$  & $4.45\pm0.31\pm0.29$\\
   3050.0 & $91\pm10$ & 2 & 59.11  & 14.90 &$10.1\pm1.1\pm0.6$ &  $3.29\pm0.17\pm0.09$\\
   3060.0 &  $78\pm9$ & 2 & 59.21  & 15.06   &  $8.5\pm1.0\pm0.6$ &  $3.03\pm0.17\pm0.10$ \\
   3080.0 &  $162\pm13$& 1  &58.97& 30.73  &    $8.9\pm0.7\pm0.5$ &  $3.11\pm0.12\pm0.08$\\
   3400.0 &  $2\pm1$ & 0   &63.34 & 1.73 & $1.8\pm1.3\pm0.4$   & $1.54\pm0.55\pm0.18$\\
   3500.0 & $5\pm2$  & 0   &63.70 & 3.61 & $2.2\pm1.0\pm0.6$ & $1.73\pm0.39\pm0.22$\\
   3550.7 & $24\pm5$ & 1& 62.23 & 18.15 &$2.0\pm0.4\pm0.6$ & $1.67\pm0.17\pm0.23$ \\
   3600.2 & $14\pm4$& 1    &62.24 &9.55 &  $2.2\pm0.6\pm0.9$ & $1.78\pm0.25\pm0.35$\\
   3650.0 &  $36\pm6$& 4 & 61.20 & 48.82 & $1.1\pm0.2\pm0.1$  & $1.26\pm0.11\pm0.07$ \\
   3671.0 & $6\pm2$ &0 &  51.17& 4.59 &  $2.2\pm0.9\pm0.8$ & $1.84\pm0.37\pm0.33$\\
  \hline
  \hline
\end{tabular}
\end{center}
    \label{result1}
\end{table*}

Several sources of systematic uncertainties are considered in the measurement of the Born cross sections
and the corresponding effective FFs,
including those of tracking, PID,
$E/p$ requirement, background estimation, theory uncertainty
from radiative corrections, FF model dependence
and integrated luminosity.

{\it (a) Tracking and PID :} The uncertainties of tracking and PID efficiencies for proton/antiproton
are investigated using almost background-free control samples
$J/\psi\rightarrow p\bar{p}\pi^{+}\pi^{-}$ and $\psi(3686)\rightarrow \pi^{+}\pi^{-}J/\psi\rightarrow \pi^{+}\pi^{-}p\bar{p}$.
The differences of tracking and PID efficiencies between data and MC simulation is $1.0\%$ per track, respectively, and
they are taken as systematic uncertainties.
{\it (b) $E/p$ requirement :}
The uncertainty of the $E/p$ requirement is also estimated using the $J/\psi\rightarrow p\bar{p}\pi^{+}\pi^{-}$ control sample. The difference between data and MC in efficiency is found to be 1.0\%
applying the same $E/p$ criteria on the proton sample, and is taken as a systematic uncertainty.
{\it (c) Background estimation :}
In the analysis, the background contamination is estimated by the MC samples. An alternative method, 2-dimensional sidebands in the proton momentum versus antiproton momentum space, is applied
to estimate the background contamination, and the difference is taken as the systematic uncertainty.
The proton/antiproton momentum sideband region is defined by $6~\sigma_{p}<|p_\text{mea}-p_\text{exp}|<11~\sigma_{p}$,
where $p_\text{exp}$ and $\sigma_{p}$ are the expected momentum and resolution of proton/antiproton
at a given c.m.~energy.
{\it (d) Radiative correction :}
In the nominal results, the radiative correction factors are estimated with the {\sc Conexc} generator.
An alternative generator, {\sc Phokhara}, is used to evaluate the theoretical calculation of
the radiative correction factors, and the differences in the resulting products $\varepsilon'$ of detection efficiency
and radiative correction factor are taken as the systematic uncertainty.
{\it (e) FFs model dependence :}
For those c.m.~energies with measured $|G_{E}/G_{M}|$ ratios, the uncertainties on the
detection efficiencies are estimated by varying the $|G_{E}/G_{M}|$ ratios with 1 standard
deviation measured in this analysis.  These systematic uncertainties are found to be less than 5.0\%.
For other c.m.~energy points, whose $|G_{E}/G_{M}|$ ratios are unknown, the uncertainties on
the detection efficiencies are evaluated to be half of the differences between the detection efficiencies
with setting $|G_{E}|=0$ or $|G_{M}|=0$, respectively, which give larger uncertainties exceeding 10.0\%.
{\it (f) Integrated luminosity :}
The integrated luminosity is measured by analyzing large-angle Bhabha scattering process,
and achieves 1.0\% in precision.

All systematic uncertainties are summarized in Table~\ref{uncertainty1}.
The total systematic uncertainty of the Born cross section is obtained by summing the individual
contributions in quadrature.
The effective FF $|G|$ is proportional to the square root of the Born cross section, and its
systematic uncertainty is half of that of the Born cross section.

\begin{table*}[htbp]
\caption{Summary of systematic uncertainties (in \%) for the Born cross sections $\sigma_\text{Born}$
         and the effective form factor $|G|$ measurements.}
\begin{center}
\footnotesize
\begin{tabular}{c|ccccccccc}
  \hline
  \hline
$\sqrt{s}$ (MeV) & ~Trk.~ & PID~ & $E/p$~ & Bkg.~ & MC gen.~ & Model~ & Lum.~ & Total ($\sigma_\text{Born}$)~ & Total ($|G|$) \\
  \hline
 2232.4 & $2.0$ & $2.0$ & $1.0$& $2.6$  & $0.4$ & $1.5$  & $1.0$& $4.4$  &  $2.2$   \\
 2400.0 & $2.0$ & $2.0$ & $1.0$& $2.0$  & $1.8$ & $4.5$  & $1.0$& $6.1$  & $3.1$\\
 2800.0 & $2.0$ & $2.0$ & $1.0$& $1.9$  & $7.5$ & $10.2$ & $1.0$& $13.2$ & $6.6$ \\
 3050.0 & $2.0$ & $2.0$ & $1.0$& $2.2$  & $0.9$ &  $4.0$ & $1.0$& $5.6$  & $2.8$ \\
 3060.0 & $2.0$ & $2.0$ & $1.0$& $3.8$  & $0.1$ & $4.1$  & $1.0$& $6.4$  & $3.2$ \\
 3080.0 & $2.0$ & $2.0$ & $1.0$& $0.0$  & $0.1$ & $4.3$  & $1.0$& $5.3$  & $2.7$ \\
 3400.0 & $2.0$ & $2.0$ & $1.0$& $0.0$  & $7.8$ & $21.9$ & $1.0$& $23.5$ & $11.8$ \\
 3500.0 & $2.0$ & $2.0$ & $1.0$& $20.0$ & $7.0$ & $12.9$ & $1.0$& $25.0$ & $12.5$  \\
 3550.7 & $2.0$ & $2.0$ & $1.0$& $20.8$ & $9.0$ & $14.3$ & $1.0$& $27.0$ & $13.5$ \\
 3600.2 & $2.0$ & $2.0$ & $1.0$& $35.7$ & $4.3$ & $11.6$ & $1.0$& $37.9$ & $18.9$ \\
 3650.0 & $2.0$ & $2.0$ & $1.0$& $3.3$  & $0.9$ & $9.7$  & $1.0$& $10.8$ & $5.4$ \\
 3671.0 & $2.0$ & $2.0$ & $1.0$& $33.3$ & $0.7$ & $13.3$ & $1.0$& $36.0$ & $18.0$\\
  \hline
  \hline
\end{tabular}
    \label{uncertainty1}
\end{center}
\end{table*}

\subsection{\boldmath {Extraction of the electromagnetic $|G_{E}/G_{M}|$ ratio} \label{secd}}

\begin{figure*}[htbp]
\begin{center}
\begin{overpic}[width=0.32 \textwidth,angle=0]{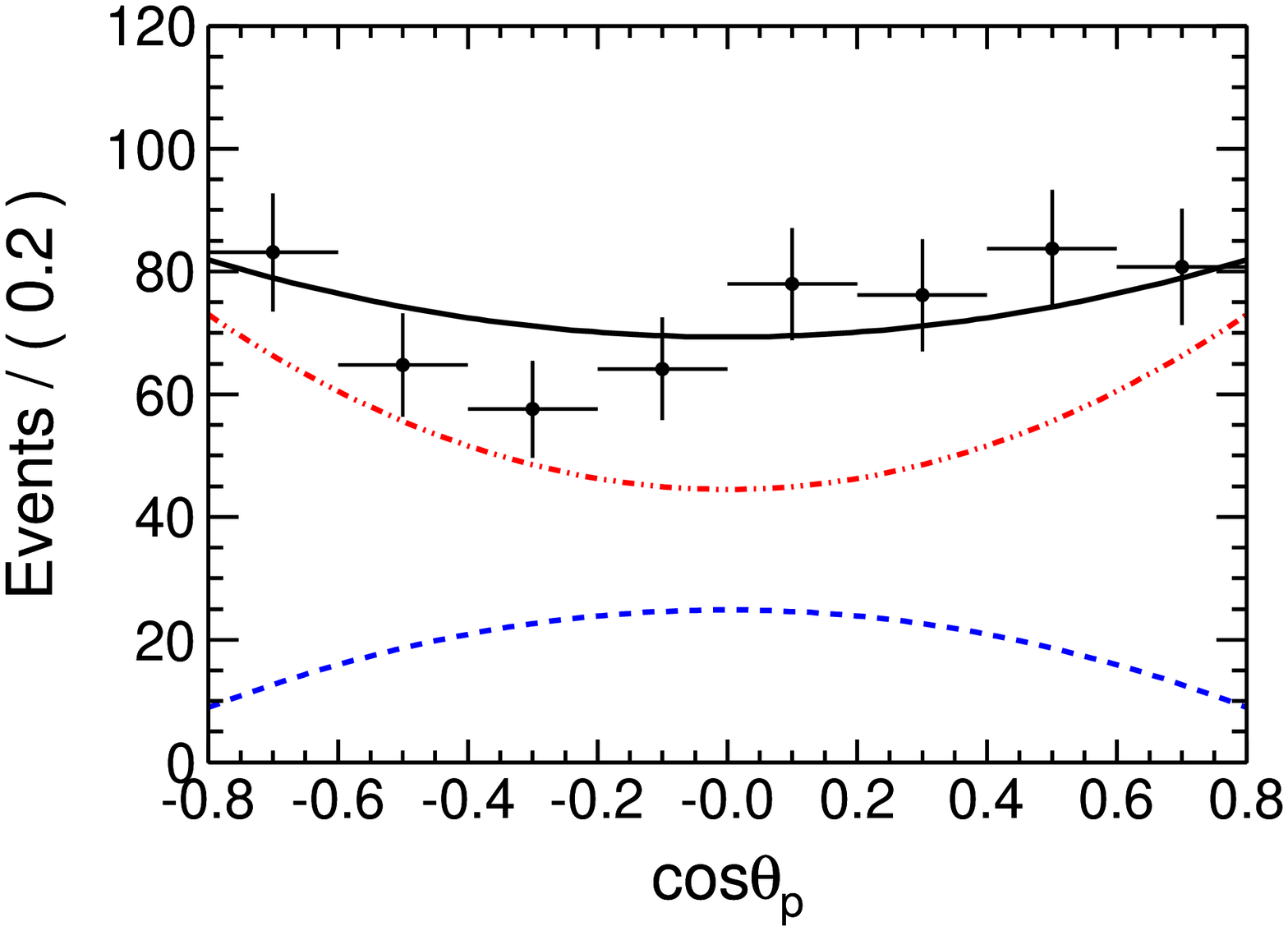}
\put(78,60){\footnotesize(a)}
\end{overpic}
\begin{overpic}[width=0.32 \textwidth,angle=0]{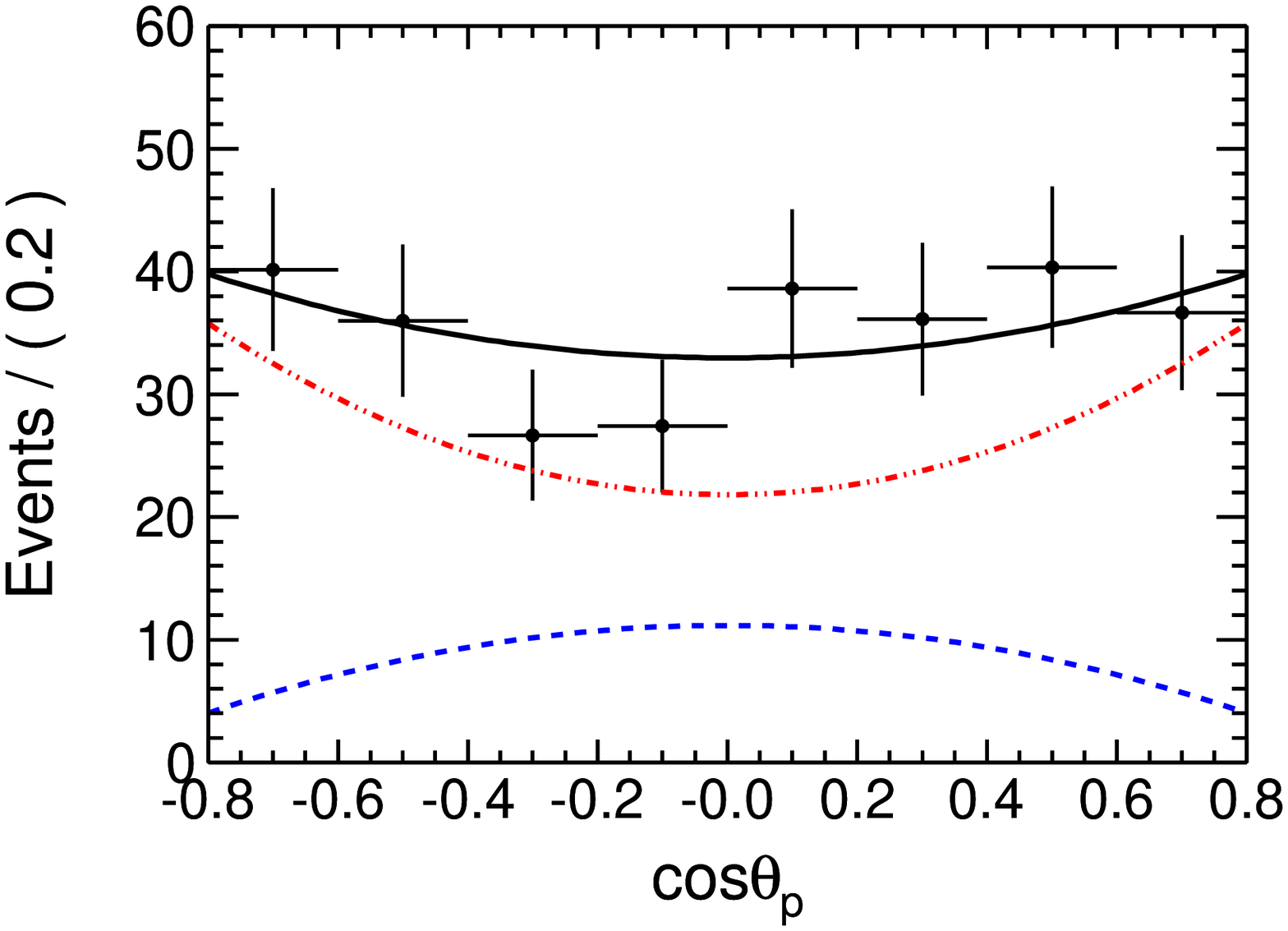}
\put(70,58){(b)}
\end{overpic}
\begin{overpic}[width=0.32 \textwidth,angle=0]{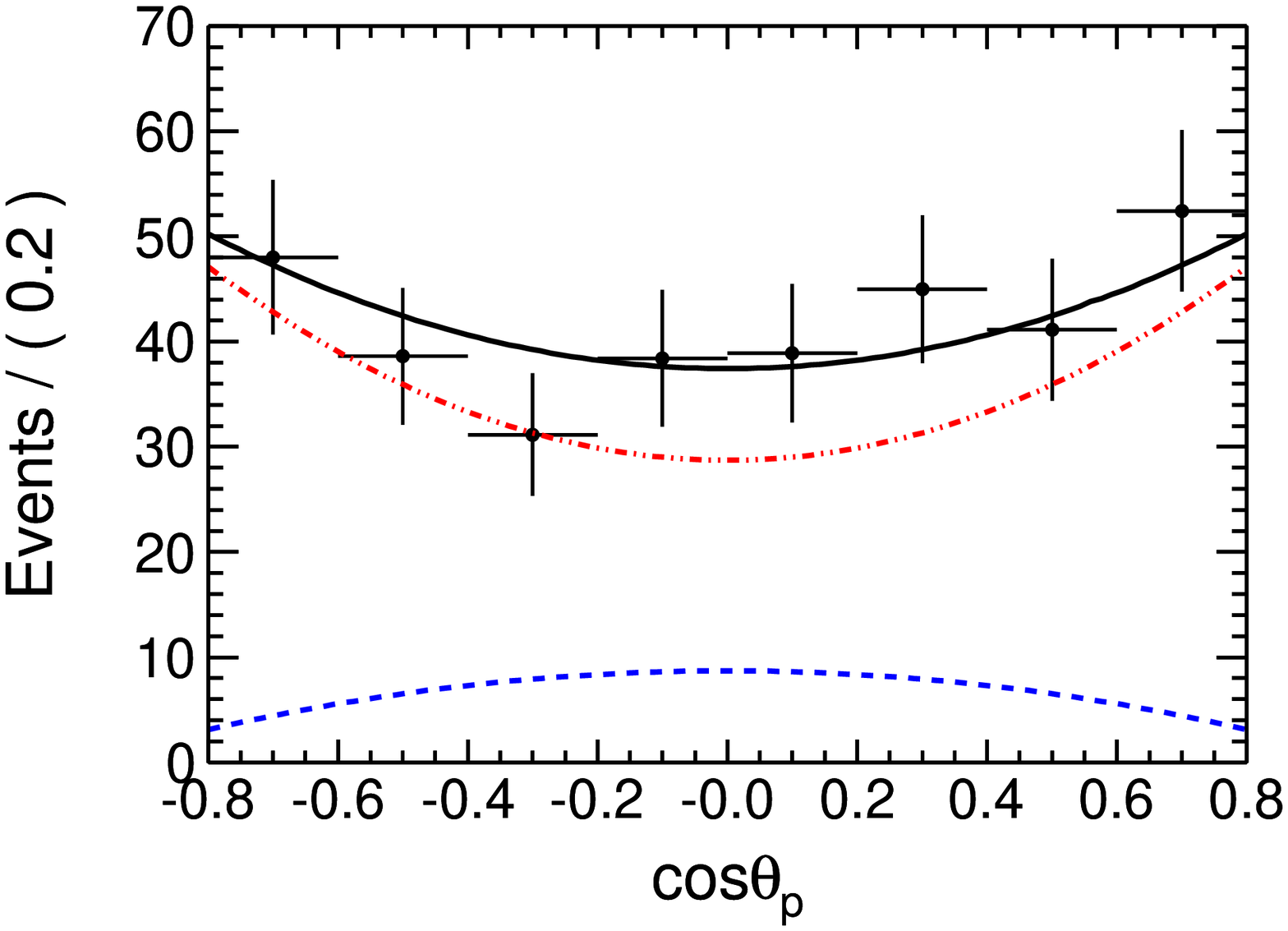}
\put(70,58){(c)}
\end{overpic}
 \caption{ Efficiency corrected distributions of $\cos\theta_p$ and fit results
 for data at c.m.~energies (a) 2232.4, (b) 2400.0 MeV and (c) a combined sample with c.m.~energy
 at 3050.0, 3060.0 and 3080.0 MeV.
 The dots with error bars represent data. The solid line (black) represents the overall fit result.
The dot-dashed line (in red) shows the contribution of the magnetic FF and the dashed line (in blue)
 of the electric FF.}
 \label{gegm}
\end{center}
\end{figure*}

\begin{figure*}[htbp]
\begin{center}
\begin{overpic}[width=4.in]{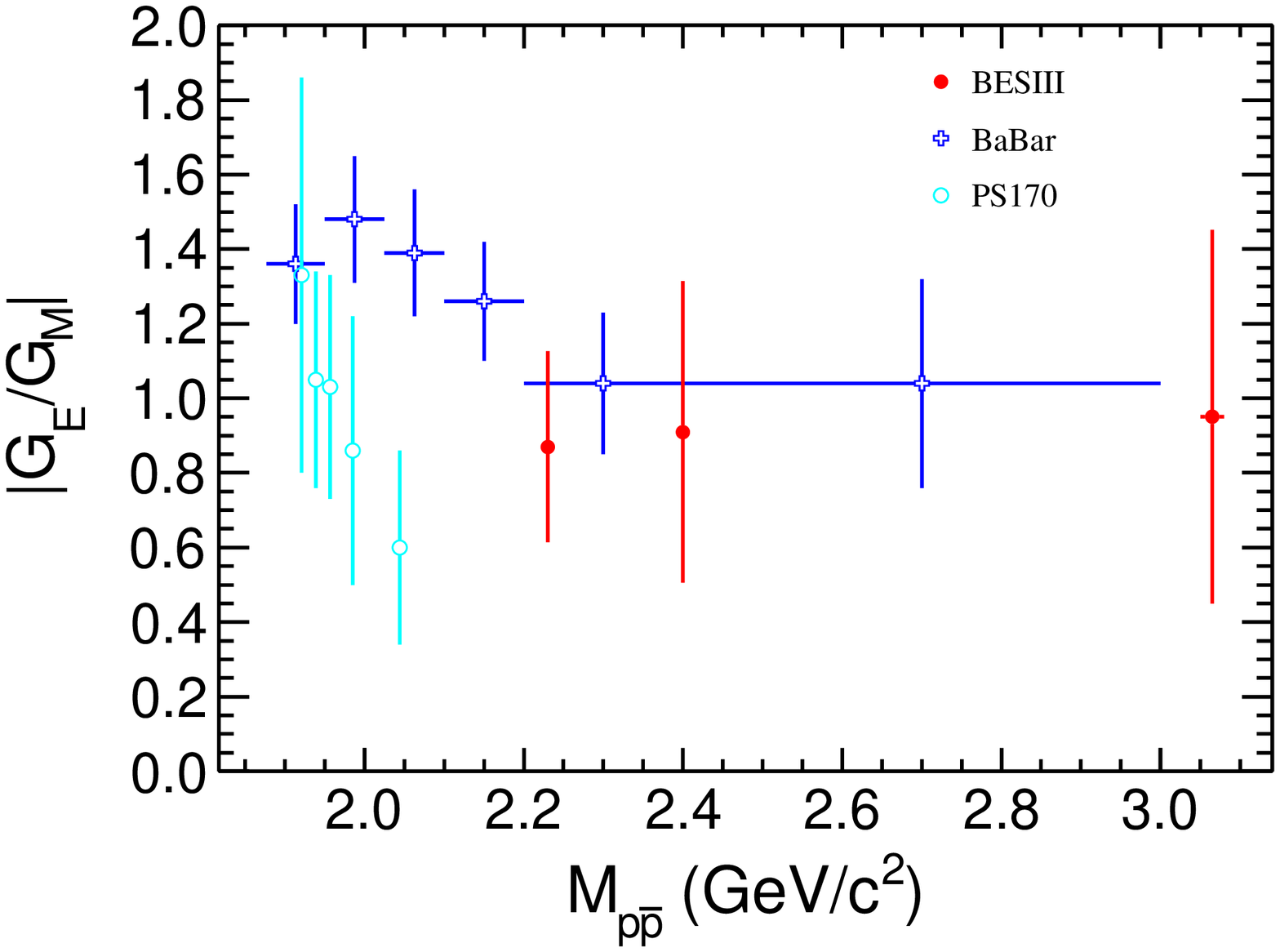}
\end{overpic}
 \caption{The measured ratio of electric to magnetic FFs $|G_E/G_M|$ at different c.m.~energy from BESIII
 (filled circles), BaBar  at SLAC (open crosses) and PS170 at LEAR/CERN (open circles). }\label{compare2}
\end{center}
\end{figure*}

The distribution of the proton polar angle $\theta_p$ depends on the electric and magnetic FFs.
The Eq.~\ref{eq1} can be rewritten as:
\begin{eqnarray}
\begin{split}
 F(\cos\theta_{p})  =&  N_\text{norm}[ 1+\cos^{2}\theta_{p}+ \\
      &\frac{4m_{p}^{2}}{s}R^{2}(1-\cos^{2}\theta_{p})],
   \label{eqx}
\end{split}
\end{eqnarray}
where $R=|G_{E}/G_{M}|$ is the ratio of electric to magnetic FFs, and
$N_\text{norm}=\frac{2\pi\alpha^{2}\beta L}{4s}[1.94+5.04\frac{m_{p}^{2}}{s}R^{2}]G_{M}(s)^2$
is the overall normalization factor.
Both $R$ and $N_{norm}$ ($G_{M}(s)$) can be extracted directly by fitting the $\cos\theta_p$ distributions
with Eq.~\ref{eqx}.
The polar angular distributions $\cos\theta_p$ are shown in Fig.~\ref{gegm} for $\sqrt{s}$ = 2232.4 and 2400.0 MeV, as well as for a
combined data sample with sub-data samples at $\sqrt{s}$ = 3050.0, 3060.0 and 3080.0 MeV. The distributions are corrected with the detection
efficiencies in different $\cos\theta_p$ bins which
are evaluated by MC simulation samples.
The distributions are fitted with Eq.~\ref{eqx}, and the fit results are also shown in Fig.~\ref{gegm}. The fit results as well as the
corresponding qualities of fit, $\chi^2/ndf$, are summarized in Table~\ref{result2}. The corresponding
ratios $R=|G_E/G_M|$ are shown in Fig.~\ref{compare2}, and the results from the previous experiments are also presented
on the same plot for comparison.

\begin{table*}[htbp]
\caption{Summary of the ratio of electric to magnetic FFs $|G_E/G_M|$, magnetic FF $|G_M|$ by fitting on the
distribution of $\cos{\theta_p}$ and \emph{method of moments} at different c.m.~energies.
For the method of fitting on $\cos\theta_{p}$, the statistical and systematic uncertainties are quoted
for $|G_{E}/G_{M}|$ and $|G_{M}|$, and the fitting quality $\chi^{2}/n.d.o.f.$ is presented.
Only statistical uncertainty is shown for the \emph{method~of~moments}.}
\begin{center}
\begin{tabular}{cccc}
\hline
\hline
\footnotesize ~~~~$\sqrt{s}$ (MeV)~~~~~~~~~& \footnotesize$~~~|G_{E}/G_{M}|~~~~~~~~~$& \footnotesize $~~~|G_{M}|$ $(\times10^{-2})~~~~~~~~$ & \footnotesize $\chi^{2}/ndf$ \\
\hline
& \multicolumn{3}{c} {\footnotesize Fit on $\cos\theta_{p}$  }  \\
\footnotesize2232.4   & \footnotesize $0.87\pm0.24\pm0.05$  &\footnotesize $18.42\pm5.09\pm0.98$  &\footnotesize 1.04\\
\footnotesize2400.0      & \footnotesize $0.91\pm0.38\pm0.12$  &\footnotesize $11.30\pm4.73\pm1.53$  &\footnotesize 0.74\\
\footnotesize (3050.0, 3080.0)&  \footnotesize$0.95\pm0.45\pm0.21$ &\footnotesize $3.61\pm1.71\pm0.82$ & \footnotesize 0.61\\
\hline
& \multicolumn{3}{c} {\footnotesize \emph{method~of~moments}  }  \\
\footnotesize 2232.4 & \footnotesize $0.83\pm0.24$  & \footnotesize $18.60\pm5.38$ & -\\
\footnotesize 2400.0 & \footnotesize $0.85\pm0.37$  &\footnotesize $11.52\pm5.01$ & -\\
\footnotesize (3050.0, 3080.0) &\footnotesize $0.88\pm0.46$  & \footnotesize $3.34\pm1.72$ & -\\
\hline
\hline
\end{tabular}
 \label{result2}
\end{center}
\end{table*}

The systematic uncertainties of the $|G_{E}/G_{M}|$ ratio and $|G_{M}|$ measurements are mainly
from background contamination, the difference of detection efficiency between data and MC, and
the different fit range of $\cos\theta_p$.
The small background contamination as listed in Table~\ref{result1} is not considered in the nominal fit. An alternative
fit with background subtraction is performed, where the background contamination is estimated by the two-dimension
sideband method, and the differences are considered as the systematic uncertainties
related to background contamination.
In the fit, the detection efficiency is evaluated with the MC simulation.
An alternative fit with corrected detection efficiency which takes into account the differences in tracking, PID and $E/p$ selection efficiency between data and MC is performed,
and the results in differences are taken as the systematic uncertainties.
Fits with ranges $[-0.8, 0.6]$ and $[-0.7, 0.7]$ in $\cos\theta_p$ are performed, and the largest differences to the nominal values are
taken as the uncertainties.
Table \ref{uncertainty2} summarizes the related systematic uncertainties for the $|G_{E}/G_{M}|$
and $|G_{M}|$ measurements.
The overall systematic uncertainties are obtained by summing all the three systematic
uncertainties in quadrature.

\begin{table*}[htbp]
\caption{Summary of systematic uncertainties (in \%) in the $|G_{E}/G_{M}|$ ratio and $|G_M|$ measurement.}
\begin{center}
\footnotesize
\begin{tabular}{c|ccc|ccc}
\hline
\hline
Source           & \multicolumn{3}{c} {$|G_{E}/G_{M}|$}  &  \multicolumn{3}{|c} {$|G_{M}|$}\\
\hline
$\sqrt{s}$ (MeV) &  $~2232.4~$&  $~2400.0~$ & $~(3050.0, 3080.0)~$ & $~2232.4~$ &  $~2400.0~$ & $~(3050.0, 3080.0)~$ \\
\hline
Background contamination & 1.1 & 7.7  & 3.2 &  1.4 & 7.7 & 3.2 \\
Detection efficiency &     2.3 & 1.1 & 4.2 &   2.3 & 1.1 & 4.2 \\
Fit range &   4.6 & 11.0 & 22.1 &      4.6  &  11.0 & 22.1  \\
\hline
Total &  5.3 &13.5 & 22.7 &  5.3 & 13.5 & 22.7  \\
\hline
\hline
\end{tabular}
    \label{uncertainty2}
\end{center}
\end{table*}

As a crosscheck, a different method, named  \emph{method of moments} (MM)~\cite{mm}, is applied
to extract the $|G_{E}/G_{M}|$ ratio, where the weighted factors in front of $G_E$ and $G_M$ may
be used to evaluate the electric or magnetic FF from moments of the angular distribution directly.
The expectation value, or moment, of $\cos^{2}\theta_{p}$, for a distribution following Eq.~\ref{eqx}
is given by:

\begin{eqnarray}
\begin{split}
  \left<\cos^{2}\theta_{p}\right> & = \frac{1}{N_\text{norm}}\int\frac{2\pi\alpha^{2}\beta C}{4s}\cos^{2}\theta_{p}[(1+\cos^{2}\theta_{p})|G_{M}|^{2} \\
 & +\frac{4m_{p}^{2}}{s}(1-\cos^{2}\theta_{p})|R^{2}|G_{M}|^{2}]d\cos\theta_{p}.
  \label{eq12}
\end{split}
\end{eqnarray}

Calculating this within the interval $[-0.8, 0.8]$ where the acceptance is non-zero and smooth,
gives for the acceptance correction:
\begin{equation}
 R=\sqrt{\frac{s}{4m^{2}_{p}}\frac{\left<\cos^{2}\theta_p\right>-0.243}{0.108-0.648\left<\cos^{2}\theta_p\right>}},
\label{eq13}
\end{equation}
and the corresponding uncertainty:
\begin{equation}
  \sigma_{R}=\frac{0.0741}{R(0.167-\left<\cos^{2}\theta\right>)^{2}}\frac{s}{4m^{2}_{p}}\sigma_{\left<\cos^{2}\theta_{p}\right>},
\end{equation}
where $\sigma_{\left<\cos^{2}\theta_{p}\right>}$ is given by
\begin{eqnarray}
\begin{split}
& \sigma_{\left<\cos^{2}\theta_{p}\right>}  =\sqrt{\frac{1}{N-1}\left[\left<\cos^{4}\theta_{p}\right>-\left<\cos^{2}\theta_{p}\right>^{2}\right]}.
\label{eq14}
\end{split}
\end{eqnarray}

In the analysis of experimental data, $\left<\cos^{2}\theta_{p}\right>$ and $\left<\cos^{4}\theta_p\right>$ are
the average of $\cos^{2}\theta_{p}$ and $\cos^{4}\theta_{p}$ which are calculated
taking the detection efficiency event-by-event into account:
\begin{equation}
\left<\cos^{2,4}\theta_{p}\right>=\overline{\cos^{2,4}\theta_{p}}=\frac{1}{N}\sum_{i=1}^{N}\cos^{2,4}\theta_{pi}/\varepsilon_i,
\end{equation}
where $\varepsilon_i$ is the detection efficiency with the $i$th event's kinematics as estimated
by the MC simulation.

The extracted $|G_{E}/G_{M}|$ ratios and $|G_{M}|$ by MM at different c.m.~energies
are also shown in Table~\ref{result2}, where $|G_{M}|$ is calculated by $N_{norm}$ in Eq.~\ref{eqx} using the measured $|G_{E}/G_{M}|$
ratio. The results are well consistent with those extracted by
fitting the distribution of polar angle $\cos\theta_p$, and the statistical uncertainty is
found to be comparable between the two different methods due to the same number of events.

\section{\boldmath SUMMARY}

Using data at 12 c.m.~energies between 2232.4 MeV and 3671.0 MeV collected with the BESIII
detector, we measure the Born cross sections of $e^{+}e^{-}\rightarrow p\bar{p}$ and
extract the corresponding effective FF $|G|$ under the assumption $|G_{E}|=|G_{M}|$.
The results are in good agreement with previous experiments.
The precision of the Born cross section with $\sqrt{s}\leq3.08$ GeV is between $6.0\%$ and $18.9\%$
which is much improved comparing with the best precision of previous results (between $9.4\%$
and $26.9\%$) from BaBar experiment~\cite{babar2};
and the precision is comparable with those of previous results at $\sqrt{s}>$ 3.08 GeV.
The $|G_{E}/G_{M}|$ ratios and $|G_{M}|$ are extracted at the c.m.~energies $\sqrt{s}=2232.4$ and
2400.0 MeV and a combined data sample with c.m.~energy of 3050.0, 3060.0 and 3080.0 MeV,
with comparable uncertainties to previous experiments.
The measured $|G_{E}/G_{M}|$ ratios are close to unity which are consistent with those of
the BaBar experiment in the same $q^{2}$ region.
At present, the precision of the $|G_{E}/G_{M}|$ ratio is dominated by statistics.
A MC simulation study shows that the precision can achieve 10\% or 3.0\%
if we have a factor of 5 or 50 times higher integrated luminosity.
In the near future, a new scan at BEPCII with c.m.~energy ranging between 2.0 GeV and 3.1 GeV
is foreseen to improve the precision of the measurement on $|G_{E}/G_{M}|$ ratio in a
wide range.

\section{ACKNOWLEDGMENTS}

The BESIII collaboration thanks the staff of BEPCII and the IHEP computing center for their strong support.
We are grateful to Henryk Czyz for providing us the new
{\sc Phokhara} generator with the scan mode. This work is supported in part by National Key
Basic Research Program of China under Contract No. 2015CB856700;
National Natural Science Foundation of China (NSFC) under Contracts Nos. 10935007, 11121092,
11125525, 11235011, 11322544, 11335008, 11375170, 11275189, 11078030, 11475164, 11005109, 11475169, 11425524;
the Chinese Academy of Sciences (CAS) Large-Scale Scientific Facility Program;
Joint Large-Scale Scientific Facility Funds of the NSFC and CAS under Contracts Nos. 11079008, 11179007, U1232201, U1332201;
CAS under Contracts Nos. KJCX2-YW-N29, KJCX2-YW-N45; 100 Talents Program of CAS; INPAC and Shanghai Key Laboratory for Particle Physics and Cosmology; German Research Foundation DFG under Contract No. Collaborative Research Center CRC-1044; Istituto Nazionale di Fisica Nucleare, Italy; Ministry of Development of Turkey under Contract No. DPT2006K-120470; Russian Foundation for Basic Research under Contract No. 14-07-91152; U. S. Department of Energy under Contracts Nos. DE-FG02-04ER41291, DE-FG02-05ER41374, DE-FG02-94ER40823, DESC0010118; U.S. National Science Foundation; University of Groningen (RuG) and the Helmholtzzentrum fuer Schwerionenforschung GmbH (GSI), Darmstadt; WCU Program of National Research Foundation of Korea under Contract No. R32-2008-000-10155-0.

\end{document}